\documentclass[intlimits,twoside,a4paper]{article}
\usepackage[cp1251]{inputenc}

\usepackage[eqsecnum]{cmpj3}

\usepackage{bm}

\issue{2026}{29}{1}{13503}
\doinumber{10.5488/CMP.29.13503}

\title[Alcohols on temperature]
{II. Temperature trends in the properties of simple monohydric alcohols.
Molecular dynamics simulations of united atom UAMI-EW model\thanks{Dedicated to the
memory of Prof. Stefan Sokolowski}
}

\author[M. Aguilar, E. N\'{u}\~{n}ez-Rojas,  O. Pizio]
{M. Aguilar\orcid{0000-0003-3850-1188}\refaddr{label1},
E. N\'{u}\~{n}ez-Rojas\orcid{0000-0002-1263-0478}\refaddr{label2},
O. Pizio\orcid{0000-0001-8333-4652}\refaddr{label1}
\thanks{Corresponding author: \email{oapizio@gmail.com}}
}

\addresses{
\addr{label1}
Instituto de de Qu\'{i}mica, Universidad Nacional Aut\'{o}noma de M\'{e}xico,
Circuito Exterior, 04510, Cd. Mx., M\'{e}xico
\addr{label2}
Departamento de Qu\'{i}mica, Universidad Aut\'{o}noma Metropolitana-Iztapalapa,
Av. San Rafael Atlixco 186, Col. Vicentina, 09340, Cd. Mx., M\'{e}xico,
}


\date{Received 11 January 2026; accepted 05 February 2026; published 30 March 2026}
\begin{document}

\maketitle

\begin{abstract}
We explore the dependence of a wide set of
properties of monohydric alcohols on temperature
by using the isobaric-isothermal molecular dynamics computer simulations.
Namely, methanol (MeOH), ethanol (EtOH) and 1-propanol (PrOH) alcohols are studied.
The recently proposed united atom, non-polarizable force field for each of alcohols
[V. Garc\'{i}a-Melgarejo et al., J. Mol. Liq., 2021, \textbf{323}, 114576]
is applied for this purpose.
Accuracy of the force field is discussed comparing predictions from simulations and
experimental data for density, dielectric constant, surface tension,
and self-diffusion coefficient.
Supplementary insights concerning applicability of the model are obtained 
by exploration of the composition dependence of various properties 
for MeOH--PrOH mixtures.
Peculiarities of mixing of species in this system are elucidated
in terms of density, excess mixing volume and excess mixing enthalpy.
Static dielectric constant of the mixture and the corresponding excess
are obtained.
Perspectives of modelling are commented finally.

\printkeywords
%
%
\end{abstract}

\section{Introduction}

Theoretical modelling of alcohols and their mixtures with water and other substances
is of much importance for chemical engineering. This topic generated enormous
amount of studies using experiments, computer simulations and semi-analytical
constructions. In spite of much efforts, the problem of describing the properties
of these systems at different pressures, temperatures and chemical composition (in the
case of multi-component mixtures) is far from being solved.
Even for rather simple molecular fluids, such as lower alcohols, and their mixtures,
the current understanding of the trends of behavior of principal properties
regarding thermodynamic conditions is not complete \cite{mathias,siepmann}.


One of the most frequently used and accurate united-atom models for simple alcohols is provided by
the TraPPE database \cite{trappe}. The coexisting vapor-liquid density, boiling temperature,
vapor pressure and critical properties are used to derive parameters of the force field.
On the other hand, the UAMI model with explicit water, denoted as UAMI-EW united atom,  
non-polarizable model
for different monohydric alcohols was constructed by considering 
the density, the static dielectric constant and surface tension as targets \cite{melgarejo}.
In contrast to TraPPE united atom modelling, the UAMI-EW force field
leads to correct miscibility values for lower alcohols with 
water as documented in \cite{melgarejo}. This scheme of parametrization
was designed at a room temperature, 298.15~K, and ambient pressure, 1 bar, thermodynamic
state. It is worth noting that the TIP4P/$\varepsilon$ water model is involved
within applied parametrization.

The principal objective of the present work is to investigate 
the temperature dependence of a set of properties
of simple monohydric alcohols (MeOH, EtOH and ProOH) by using
the UAMI-EW united atom,  non-polarizable models. Very recently, we have tested these
models with respect of its applicability at a fixed temperature, 298.15~K, 
but in a wide interval of pressure values \cite{cmp2026,jcp2025}.
Namely, in reference~\cite{cmp2026} we explored the evolution of the principal properties of the UAMI-EW
model from ambient pressure, 1~bar, up to 3 kbar, for each of three alcohols in question.
We refer to this reference as part I below.
On the other hand, in reference~\cite{jcp2025}, the main objective was to elucidate  the 
trends of behavior of the hydrogen bonds network of UAMI-EW united atom model for methanol,
along with other popular methanol models~\cite{trappe,vega-met}, up to high pressure, 6 kbar.
It is well established that water molecules form a three dimensional network
of tetrahedrally coordinated molecules, see, e.g., references~\cite{ball,franks}, whereas in
liquid methanol chains and/or branched chains can be detected. 
This leads to well pronounced differences of the topology of the hydrogen
bonds network in aqueous alcohol solutions in 
comparison with pure alcohols \cite{vrhovsek,weitkamp}.

The second part of the present work is devoted to the study of MeOH--PrOH mixtures
at different temperatures. In this way we would like to elucidate the applicability of
the present UAMI-EW modelling to alcohol mixtures. In future work from our laboratory, 
this investigation will be extended to 
secondary alcohols in the spirit of the previous, quite recent studies~\cite{mathias,siepmann}.

\section{Models and simulation details}

In general terms,  within this type of modelling, the interaction  potential
between all atoms and/or groups is assumed
as the sum of Lennard-Jones (LJ) and Coulomb contributions.
Lorentz--Berthelot combination rules are used to determine the cross parameters for
the relevant potential well depths and diameters.

Molecular dynamics computer simulations of water-methanol mixtures were performed in the
isothermal--isobaric (NPT) ensemble at a  given pressure  and temperature.
We used GROMACS software~\cite{gromacs} version 5.1.2.
The simulation box in each run was cubic, the number of molecules of the species
in all cases was fixed at 3000. 
Periodic boundary conditions were used.
Temperature and pressure controls were provided by 
the application of V-rescale thermostat and Parrinello--Rahman
barostat with $\tau_T = 0.5$~ps and $\tau_P = 2.0$~ps, respectively, the timestep was 0.002 ps.
The compressibility value  $4.5\times10^{-5}$~bar$^{-1}$ was used.

The non-bonded interactions were cut-off at 1.1 nm, whereas the long-range electrostatic interactions
were handled by the particle mesh Ewald method (fourth
order, Fourier spacing equal to 0.12) with the precision $10^{-5}$.
The van der Waals correction terms to the energy and pressure were applied.
In order to maintain the geometry of alcohol
intra-molecular bonds rigid, the LINCS algorithm was used.
All the parameters for the force fields are given 
in the supplementary material to reference~\cite{melgarejo}.

After preprocessing and equilibration, consecutive simulation runs,
each with time duration not less than 10 ns,  with
the starting configuration being the last configuration from the previous
run, were performed to obtain the trajectories for the data analysis.
The results for the majority of  properties  were obtained by averaging
over 7--10 production runs.
The dielectric constant and self-diffusion coefficients were evaluated from the
entire trajectory.

\section{Results: MeOH, EtOH and PrOH alcohols}

Each of the properties below is considered for three alcohols, MeOH, EtOH and PrOH,
in order to capture the effect of length of the hydrophobic part of alcohol molecule.
In the majority of cases we intended to cover the entire interval of temperatures 
where the experimental data are available.

\subsection{Density}

\begin{figure}[h]
\begin{center}
\includegraphics[width=6.0cm,clip]{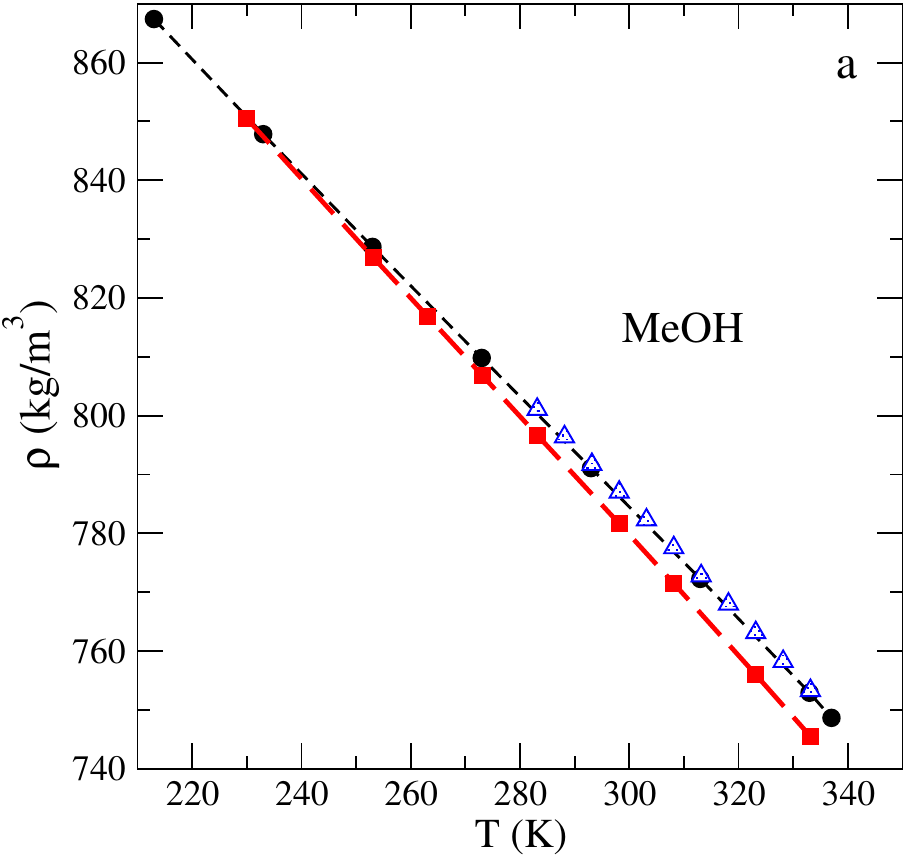}
\includegraphics[width=6.0cm,clip]{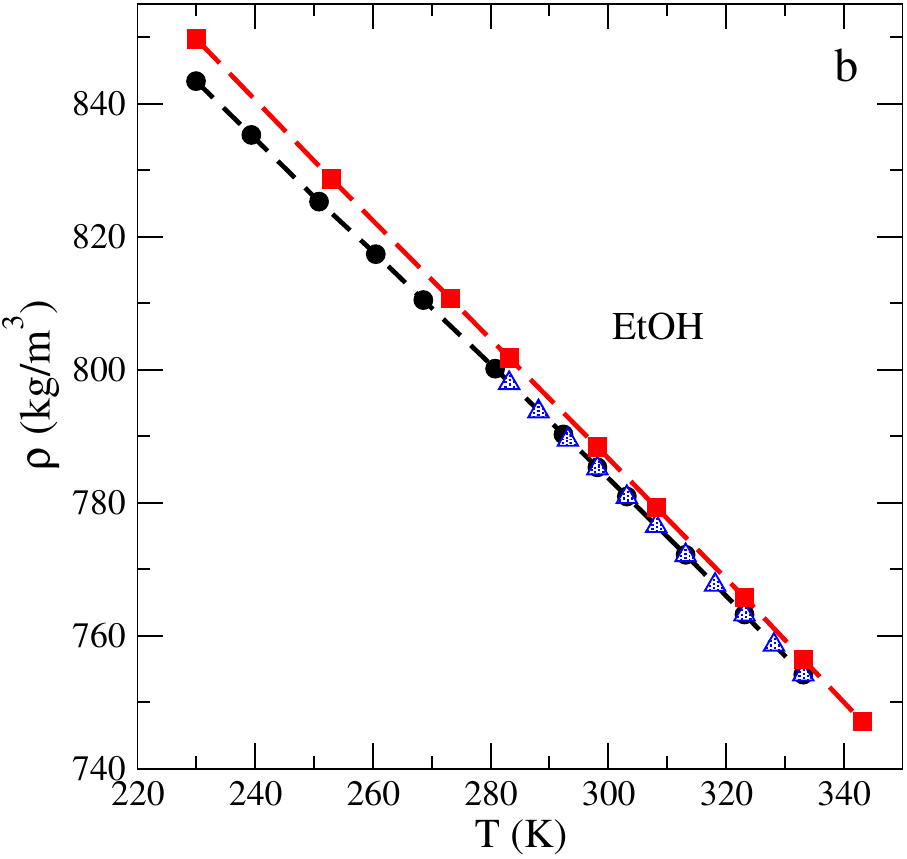}\\
\includegraphics[width=6.0cm,clip]{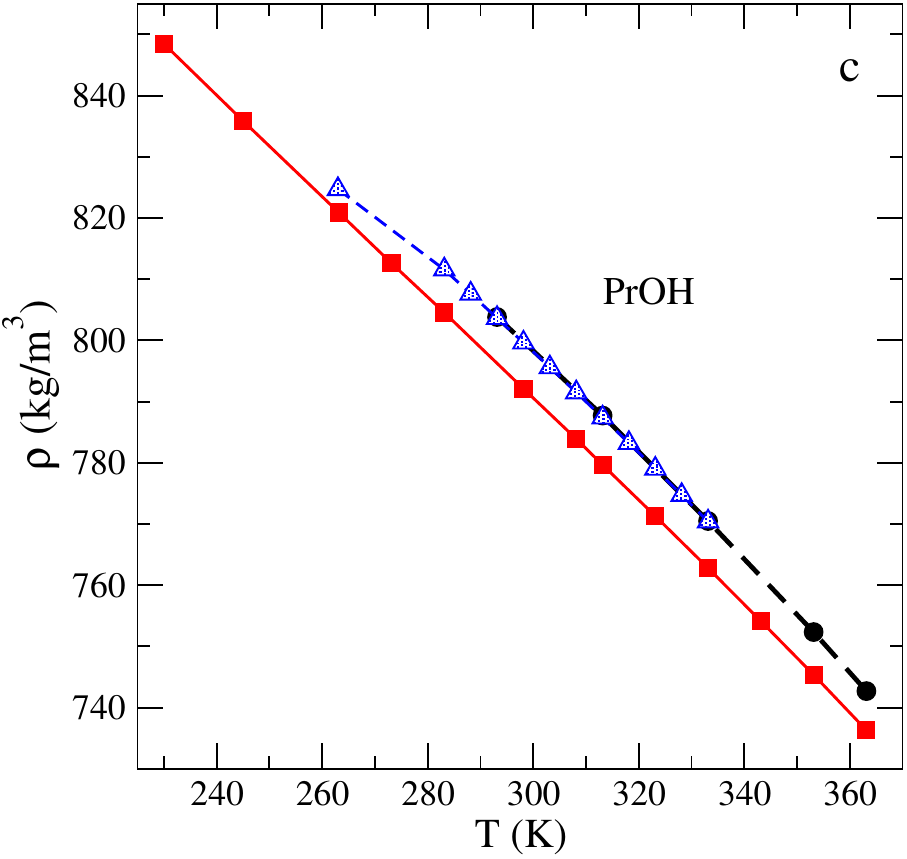}
\end{center}
\caption{(Colour online) 
Panel a: methanol density on temperature at pressure 1 bar.
The experimental data (black short-dashed line and solid circles) 
are from NIST Chemistry Webbook~\cite{nist} and from 
reference~\cite{valtz} (blue hollow triangles).  
Panel b: ethanol density on temperature. The experimental data
are from: Sun et al.~\cite{sun} (black circles) and 
reference~\cite{valtz} (blue hollow triangles).
Panel~c:~experimental data from \cite{moreau} (black circles), 
\cite{valtz} (blue hollow triangles).
Our simulation
data of UAM-EW model are given by red dashed line with
solid squares in all panels.
}
\label{fig-1}
\protect
\end{figure}

We begin with the dependence of density for each of alcohols on temperature at 
ambient pressure, 1~bar. The corresponding lines are shown in three panels of figure~\ref{fig-1}. 
The simulation results for UAMI-EW model are given by red lines decorated by solid
squares. Another notation is explained in the figure caption. We observe that the model in
question provides a quite good description of density on temperature as it follows
from comparison with experimental data. The accuracy of the model is better for MeOH and EtOH
compared to PrOH. Nevertheless, even in this case, the deviation of simulation data from 
experimental points is very small in percentage.

We would like to recall that the dependence of density on pressure at room
temperature, 298.15~K, for each of alcohols in question is described 
rather well, cf. figure~1 of reference~\cite{cmp2026}. Thus, concerning volumetric  properties
determined mainly by density, the UAMI-EW model can be used in a large
portion of $T, P$ plane with sufficient confidence.

\subsection{Dielectric constant}

The dielectric constant is one of the most important physico-chemical
properties of a given liquid, since to much extent it determines the 
properties of mixing with other substances. The static dielectric constant has served
as a target property in the parametrization of UAMI-EW model at $T = 298.25$~K
and $P = 1$~bar, besides the density and surface tension.
Our results for $\varepsilon$  at different temperatures follow
from the time-average of the fluctuations of the total
dipole moment of the system as common~\cite{martin},

\begin{equation}
\varepsilon=1+\frac{4\piup}{3{k_{\rm B}}TV}\big(\langle\bf M^2\rangle-\langle\bf M\rangle^2\big),
\end{equation}
where $k_{\rm B}$ is the Boltzmann constant and $V$ is the simulation cell volume.

\begin{figure}[h]
	\begin{center}
		\includegraphics[width=5.5cm,clip]{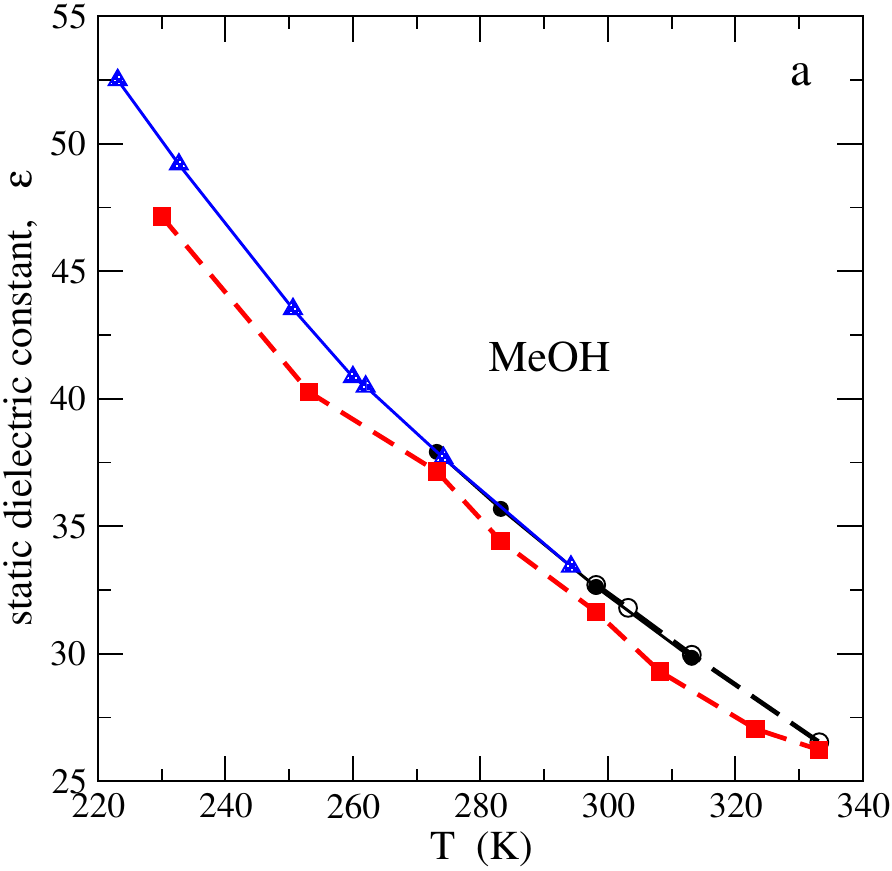}
		\includegraphics[width=5.5cm,clip]{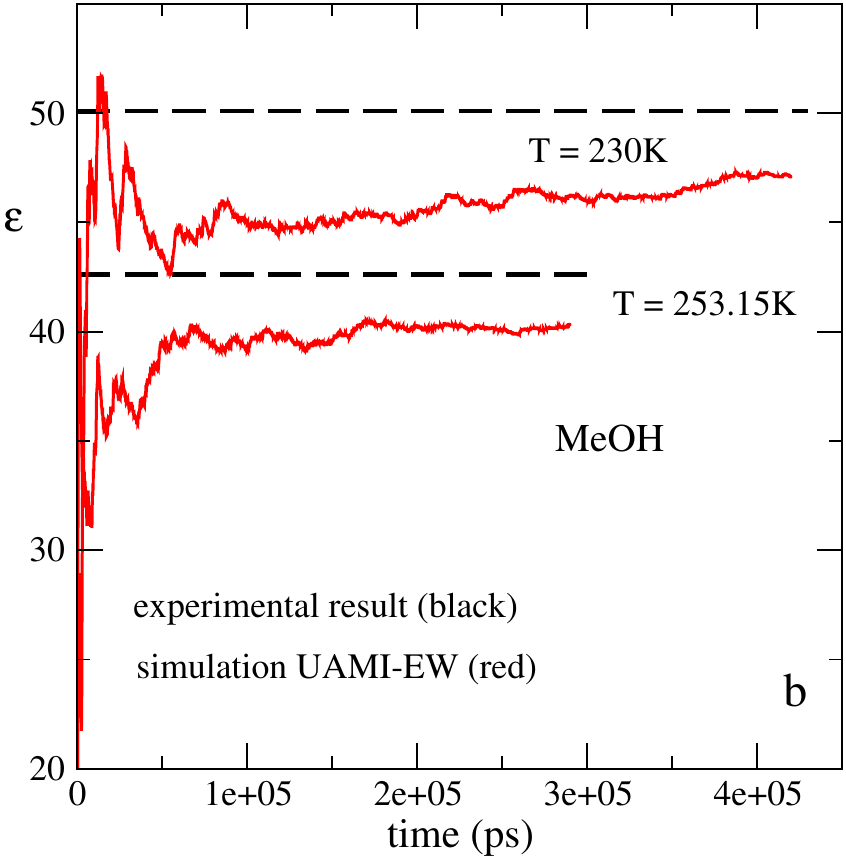} \\
		\includegraphics[width=5.5cm,clip]{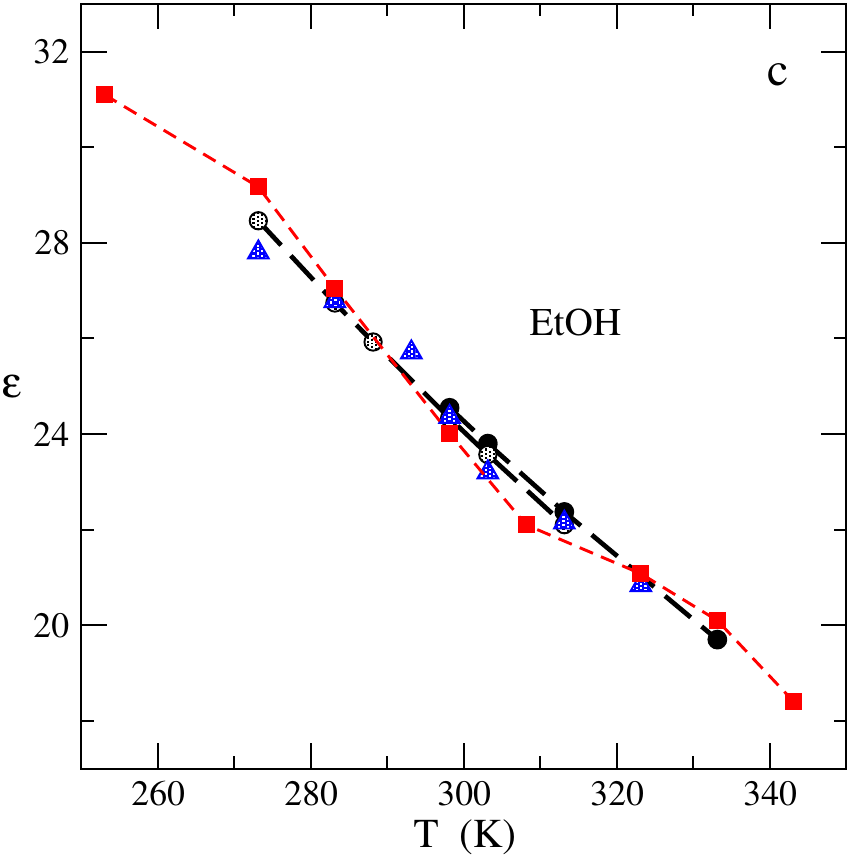}
		\includegraphics[width=5.5cm,clip]{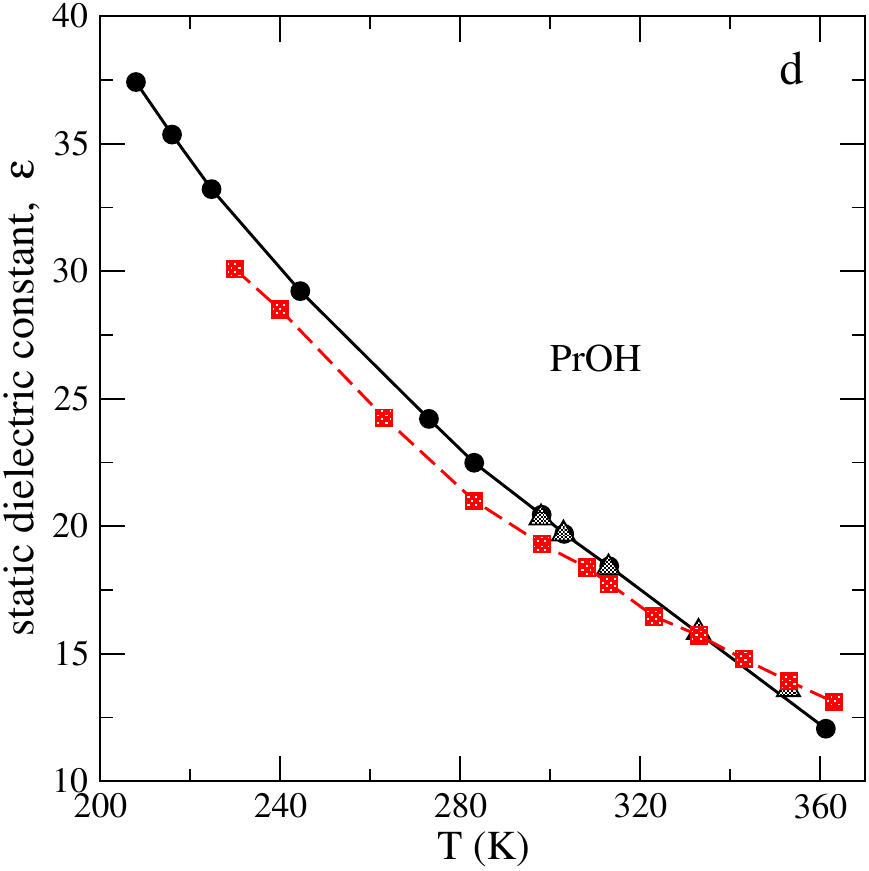}
	\end{center}
	\caption{(Colour online) Panel a:  dielectric constant of methanol on temperature. 
		The experimental data in panel a are from reference~\cite{kay} (black
		solid cicles), \cite{dannhauser} (hollow circles), \cite{davidson} (blue triangles).
		The simulation results (red squares in panels a, c and d) are for  UAMI-EW united atom  model.	
		Panel~b:~illustration of the calculations of
		dielectric constant of methanol.
		Panel~c:~experimental data are from reference~\cite{dannhauser}~(black solid circles),~\cite{kay} (shaded circles),~\cite{moriyoshi}
		(blue triangles), respectively.
		Panel~d:~experimental data are from references~\cite{moriyoshi}
		(black circles) and~\cite{dannhauser} (triangles), respectively.
	}
	\label{fig-2}
	\protect
\end{figure}

All the results for three alcohols, MeOH, EtOH and PrOH, are given in panels a, c and d
of figure~\ref{fig-2}. We keep the same nomenclature for simulation results (red lines with solid squares)
for the sake of convenience of the reader. In panel b, the time dependence of the 
value for the dielectric constant for MeOH is illustrated. The density
of each alcohol increases upon decreasing temperature (cf. figure~\ref{fig-1}), apparently longer 
in time trajectories are necessary to reach the plateau that determines the value for $\varepsilon$.
We  observe that the model for MeOH slightly underestimates the dielectric constant in
comparison with experiment (figure~\ref{fig-2}a). At low temperatures, the deviation of simulations results
from experiments becomes larger, in comparison with high temperatures where thermal fluctuations
are stronger.
For alcohols with longer hydrophobic part of the molecule, the agreement between
simulation results and experimental data is satisfactory. Hence, parametrization of the model
at $T = 298.15$~K, provides a reasonable description of the static dielectric constant in a
quite large interval of temperature changes.  This conclusion may be considered together
with the findings for $\varepsilon(P)$, cf. figure~4 of reference~\cite{cmp2026}. Both of them support a 
successful applicability of UAMI-EW model for the description of dielectric constant
of each of three alcohols in question in  the $T, P$ plane.  One of the missing elements, or
say possible extension necessary to complete,  is in the study of dielectric relaxation phenomena
for alcohols with a different number of carbon groups in the hydrophobic part of the
molecule. We hope to address this issue in our future work.

\subsection{Surface tension}

\begin{figure}[h]
\begin{center}
\includegraphics[width=6.0cm,clip]{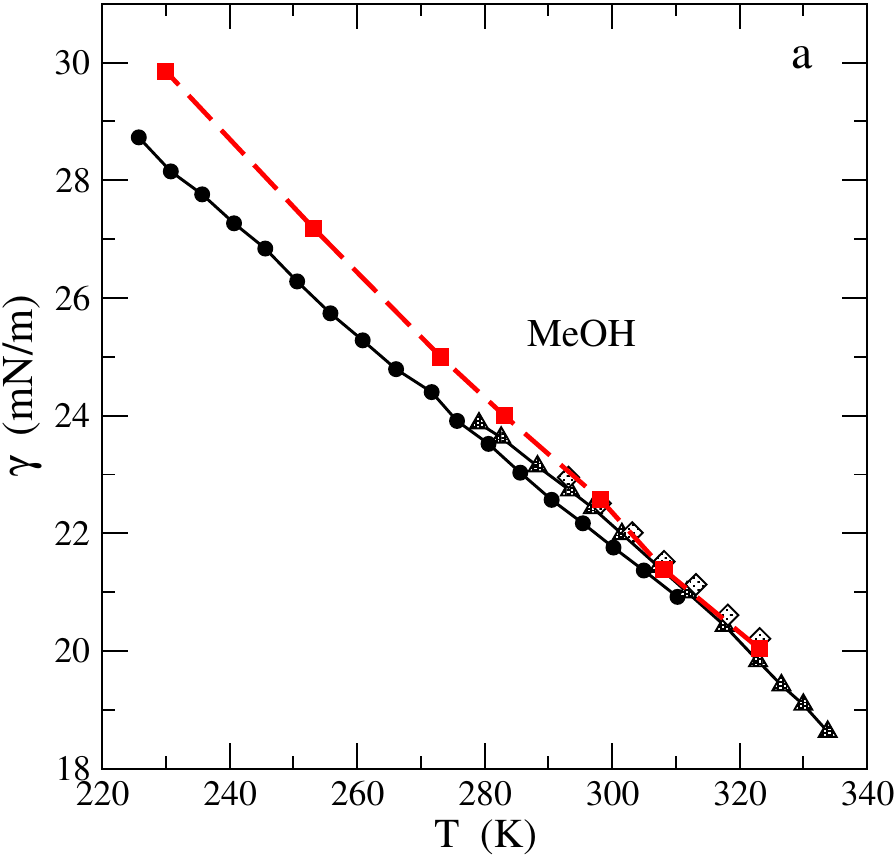}
\includegraphics[width=6.0cm,clip]{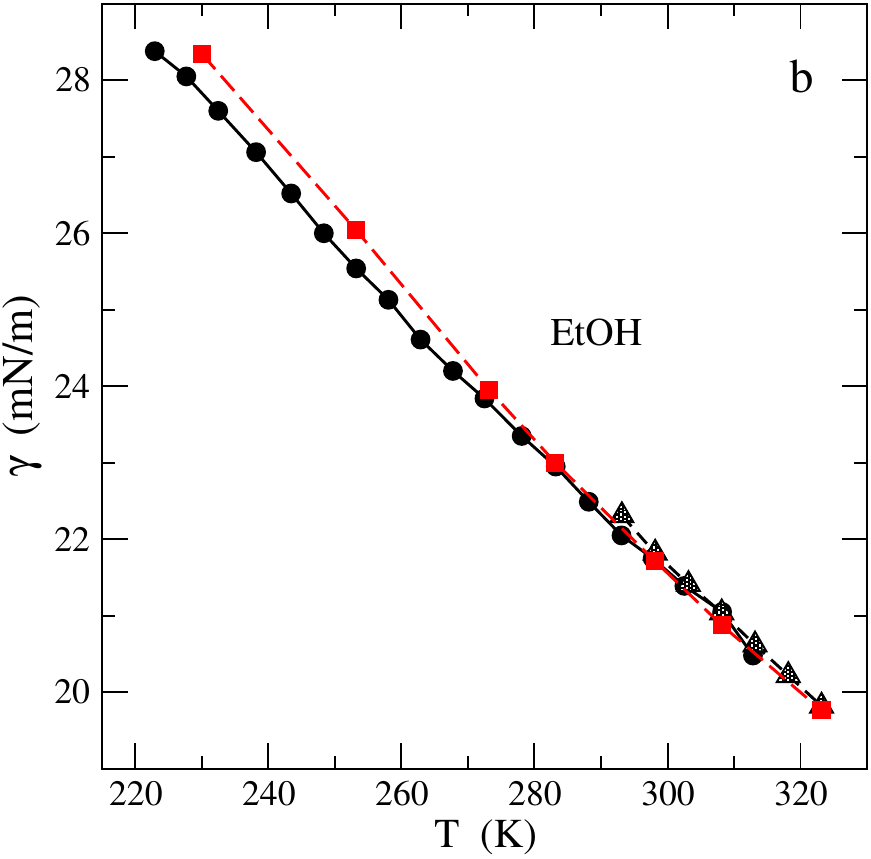} \\
\includegraphics[width=6.0cm,clip]{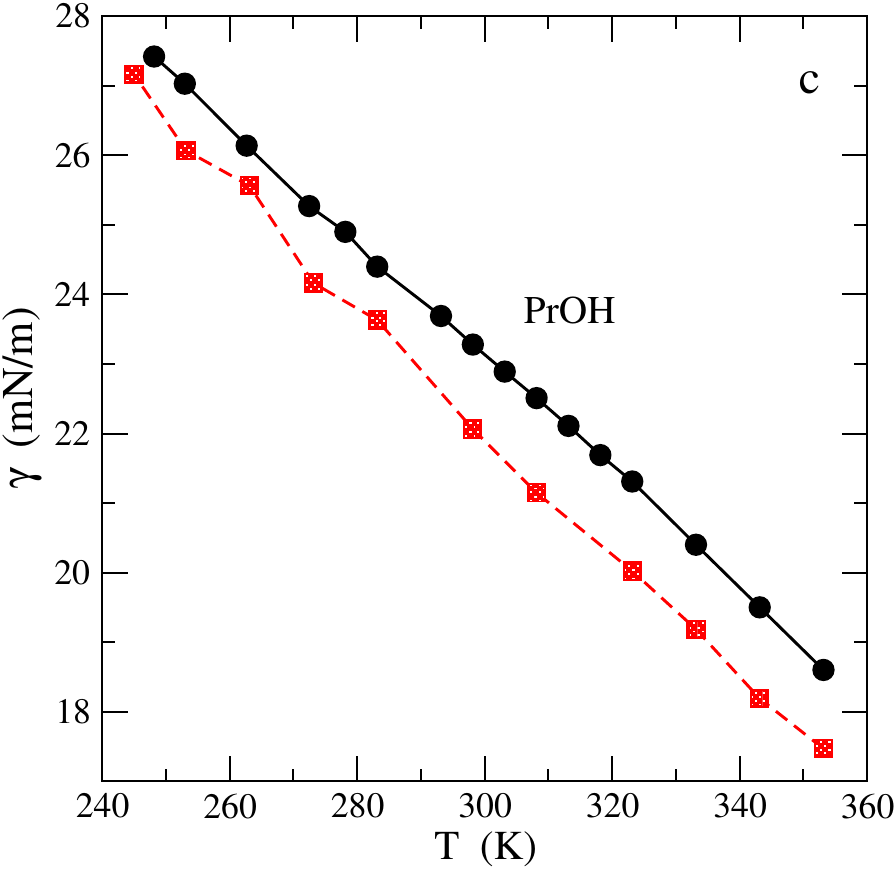}
\end{center}
\caption{(Colour online)
Panel a:  surface tension of methanol on temperature.
The experimental data in panel a are from reference\cite{strey} (circles),~\cite{souckova} (triangles) and~\cite{vazquez}
(diamonds), respectively.
Panels b and c: the same as in panel a, but for EtOH and PrOH, respectively.
Experimental data are from reference~\cite{strey} (circles --- in panels b and c) and
from~\cite{vazquez} (triangles) in panel b.
The simulation results in all panels are for UAMI-EW united atom  model (squares).
}
\label{fig-3}
\protect
\end{figure}

We proceed now to the surface tension, $\gamma$, on temperature trends. This property
was used as one of the targets in the construction of UAMI-EW model. The entire set
of results obtained is given in figure~\ref{fig-3}. 

In general terms, the UAMI-EW model correctly describes
$\gamma(T)$ dependencies for each of three alcohols. An overall agreement 
between simulation results and experimental data is observed. This agreement is better
for MeOH and EtOH (panels a and b). Nevertheless, at low temperatures, the UAMI-EW model overestimates
the surface tension values, in comparison with experimental results. On the other hand,
for PrOH, the model underestimates the values for the surface tension in the
entire temperature interval under study (panel c). The absolute value of deviation
of simulation results from experiment is around 5\%.
This behavior reflects the impossibility to reproduce several properties simultaneously
with high precision at the level of united atom modelling. Still, the optimum balance of accuracy 
with small margins for a set of properties can be reached.

We have evaluated the density profiles of alcohol molecules across liquid--vapor
interface, but do not show them for the sake of brevity. It may be of interest to
discuss them more in detail in a separate publication.

\subsection{Self-diffusion coefficient}

Evaluation of the quality of the model often involves results for the
self-diffusion coefficient, $D$. Note that $D$ was not considered as a target
property within the multi-step parametrization of the UAMI-EW model~\cite{melgarejo}.

\begin{figure}[h]
\begin{center}
\includegraphics[width=6.0cm,clip]{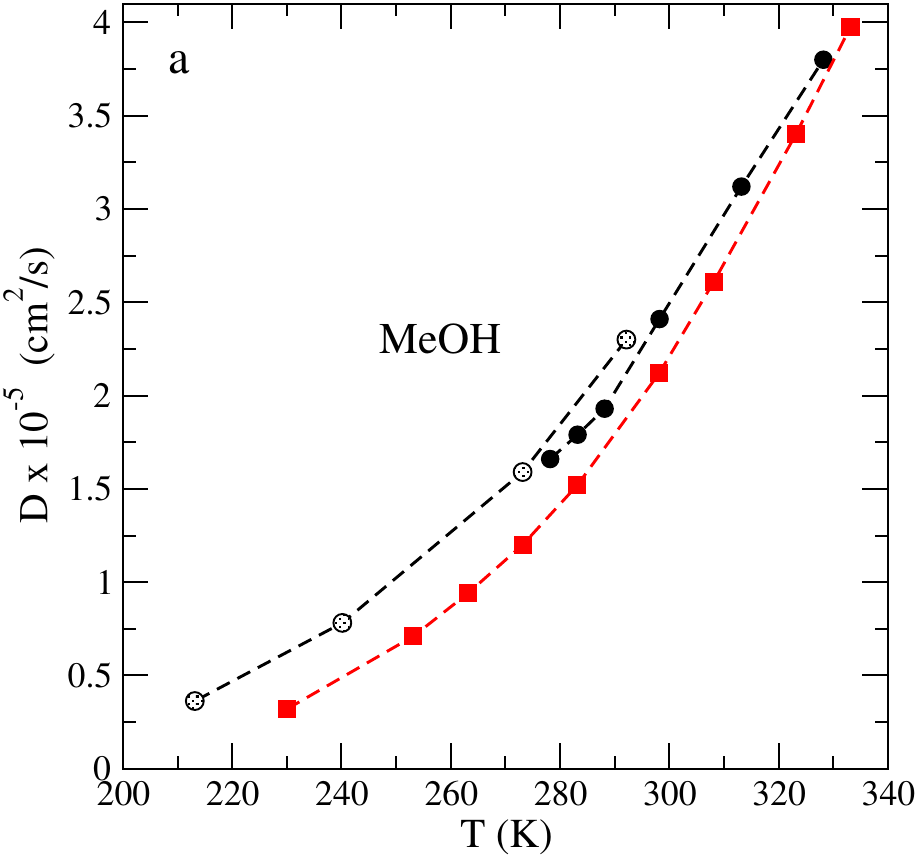}
\includegraphics[width=6.0cm,clip]{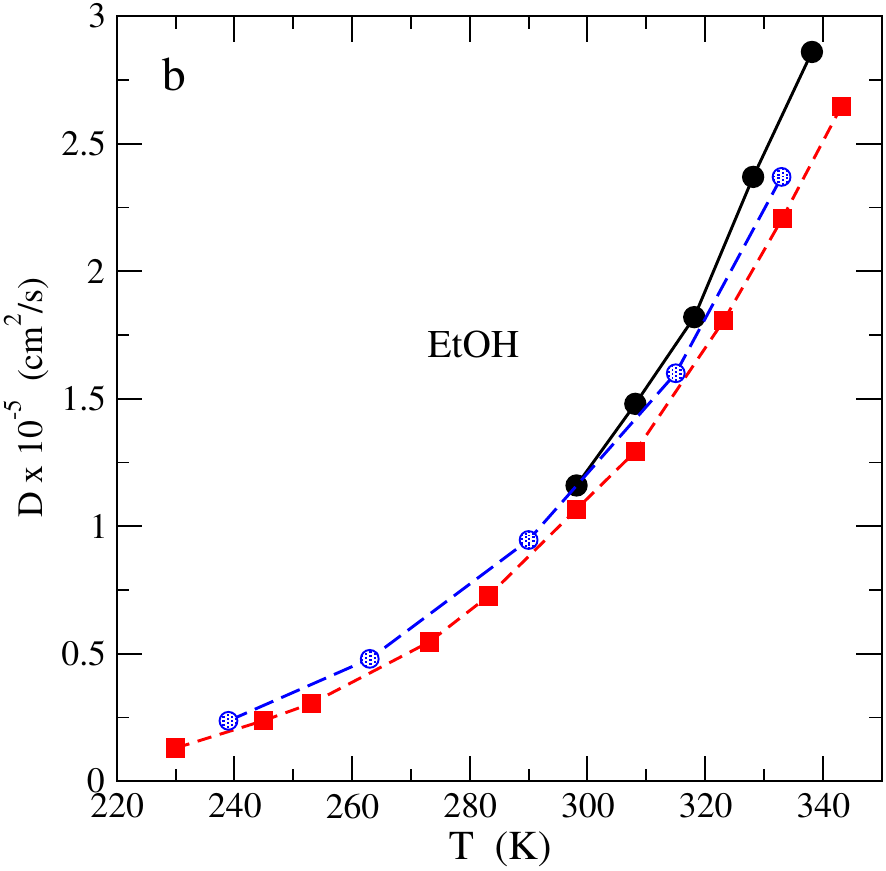}
\includegraphics[width=6.0cm,clip]{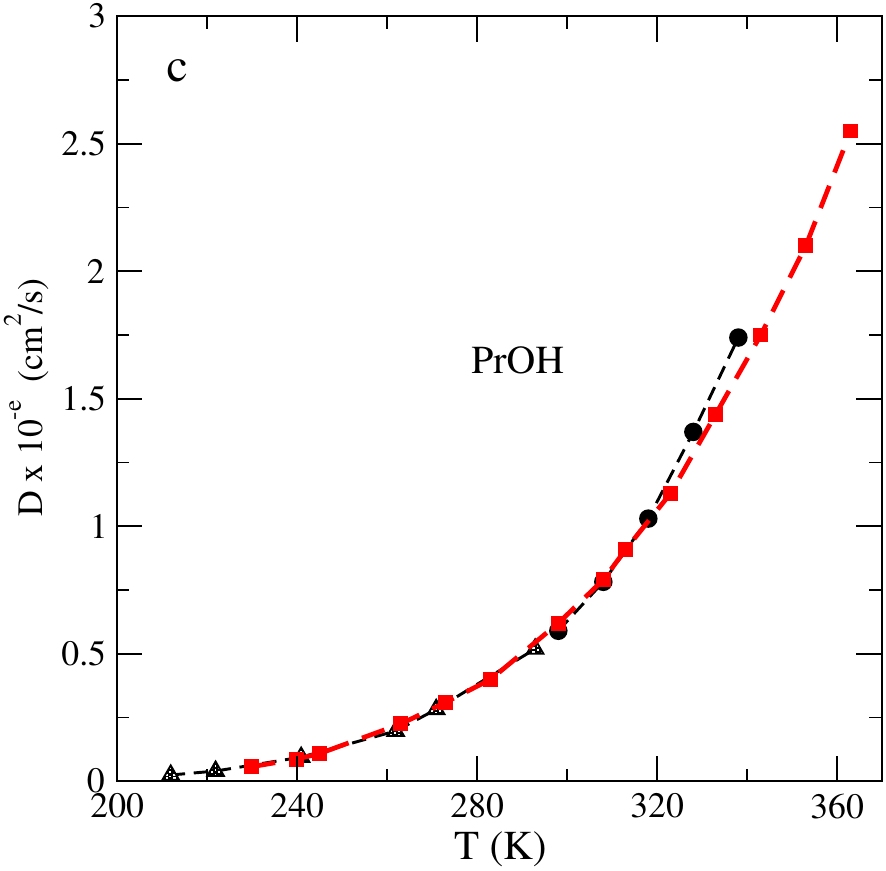}
\end{center}
\caption{(Colour online)
Panel a: self-diffusion coefficient of methanol on temperature. 
Experimental data are from \cite{hurle,karger} (solid circles),
and \cite{karger} (hollow circles).
Panel b: self diffusion coefficient of ethanol on 
temperature (experimental data --- \cite{pratt}, \cite{karger}, solid and hollow
circles, respectively).
Panel~c:~self diffusion coefficient of PrOH on temperature 
(experimental data --- \cite{pratt,shaker},
circles and triangles, respectively).
}
\label{fig-4}
\protect
\end{figure}

One of the common routes to obtain $D$,  is from the
mean square displacement of particles. On the other hand, it may be also calculated from the
velocity
auto-correlation function.
We calculate $D$  by the former method, via the Einstein relation,
\begin{equation}
D =\frac{1}{6} \lim_{t \rightarrow \infty} \frac{\rd}{\rd t} \vert {\bf r}(\tau+t)
-{\bf r}(\tau)\vert ^2,
\end{equation}
where  $\tau$ denotes the time origin. Default settings of GROMACS were used
for the separation of the time origins.

The dependence of $D$ on temperature from simulations of three alcohols in question
is shown in figure~\ref{fig-4}. Besides, the figure contains available experimental data
for sake of comparison. It can be seen that $D$ increases with increasing
temperature for each alcohol, in agreement with the trends of experimental data.
Thermal movement facilitates self-diffusion, as expected.
The values of $D$ for MeOH from simulations are underestimated over the entire temperature range 
compared with the experimental results. On the other hand, the agreement between simulations
and experiments is very good for EtOH and PrOH.
The temperature interval is limited from above by the boiling temperature for each alcohol.
At low temperatures, we have not attempted to explore the behavior of self-diffusion coefficient
very close to solidification.  Having in mind a satisfactory performance of the UAMI-EW force field 
for $D(P)$  at $T = 298.15$~K, cf. figure~5 of reference~\cite{cmp2026}, we would just like to note
that the model provides a successful description of the self-diffusion coefficient for 
liquid alcohols under study in the ($T$, $P$)-plane. More extensive insights into the dynamic properties
can be obtained by performing additional simulations and analysis 
of the resulting different auto-correlation functions.

On the other hand, it is worth to comment that certain insights into
the structural properties of primary alcohols within UAMI-EW model were discussed in
the recent work from this laboratory~\cite{bermudez}. 
Principally, they refer, however, to various distribution functions and 
coordination numbers of alcohol--water mixtures. For pure alcohols, our very recent research 
on the microscopic structure restricted to the behavior of methanol solely, in a quite wide 
interval of values for pressure~\cite{jcp2025}. A reasonable agreement of the
structural properties of methanol model, in comparison with the X-ray and neutron scattering 
results for the structure factors, was obtained. Moreover, cooperative effects
of the hydrogen bonds network in MeOH were elucidated. We expect that progress in the
structure measurements for other primary alcohols at different pressures and temperatures
will be reached soon, and will make possible a comparison of simulation results 
for the structure and experimental data. 

Natural extension of the modelling of alcohols explored above is in its application 
to mixtures. Specifically, we would like to investigate whether the level of accuracy of the results
for pure alcohols is sufficient to provide appropriate and correct trends 
of the behavior for their mixing properties.  

\section{Application of the UAMI-EW model to MeOH--PrOH mixtures}

Most frequent and abundant simulations were performed for mixtures 
of simple primary alcohols with water. 
On the other hand, several studies were focused on 
the description of mixing properties of various alcohol binary mixtures.
We refer to references~\cite{mathias,siepmann} as illustrative examples.
These systems are of much importance for practical applications.

Sophistication of the models for single-component alcohols, 
starting from the united atom description~\cite{trappe,melgarejo} 
up to all-atom~\cite{wensink} and polarizable models, see, e.g.,~\cite{anisimov},
resulted in their application
for binary mixtures of primary alcohols. 
Namely, we found and refer here to some of the results coming from 
all-atom modelling~\cite{madhurima} and of ab initio modelling~\cite{lone}.
These two works from the same laboratory intended to elucidate a few principal
features of the behavior of binary alcohol mixtures.
Thus, we decided to explore the mixing properties of MeOH and PrOH and extend 
the evaluation of accuracy of UAMI-EW modelling. It is worth mentioning
that binary alcohol mixtures are much less studied experimentally,
in comparison with alcohol--water solutions. Some of experimental 
observations, however, challenge computer simulations modelling.
Namely, as documented in reference~\cite{siepmann}, temperature dependence
of mixing enthalpy for mixtures of primary and secondary alcohols exhibits
peculiarities that require appropriate modelling. 

\begin{figure}[h]
\begin{center}
\includegraphics[width=6.0cm,clip]{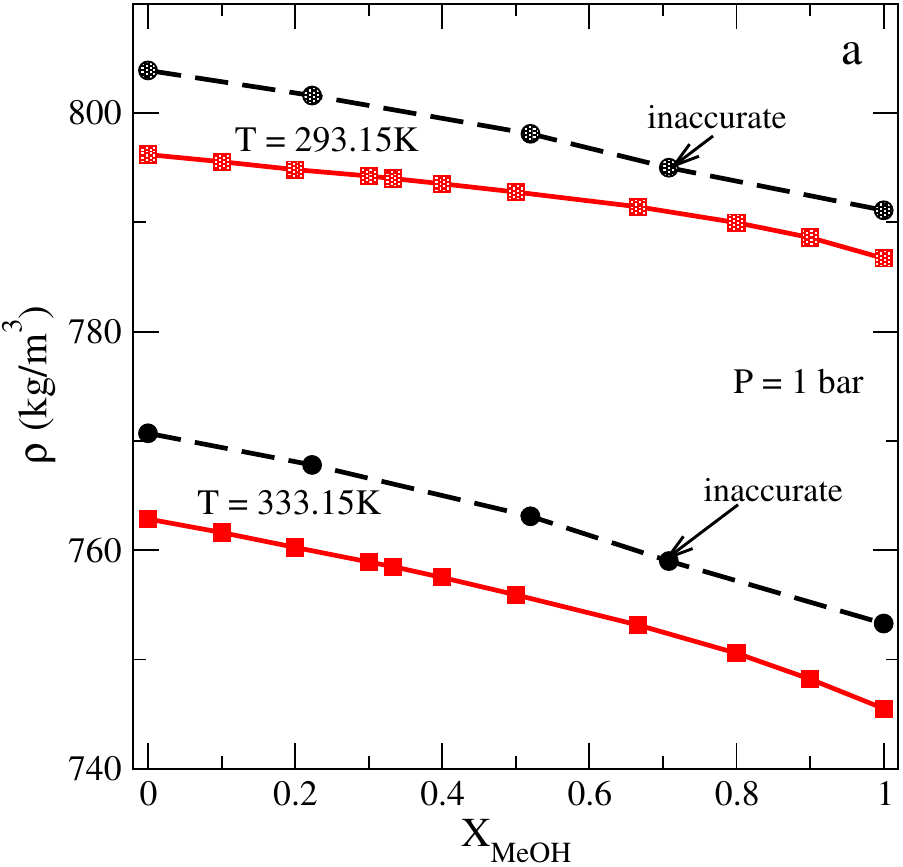}
\includegraphics[width=6.0cm,clip]{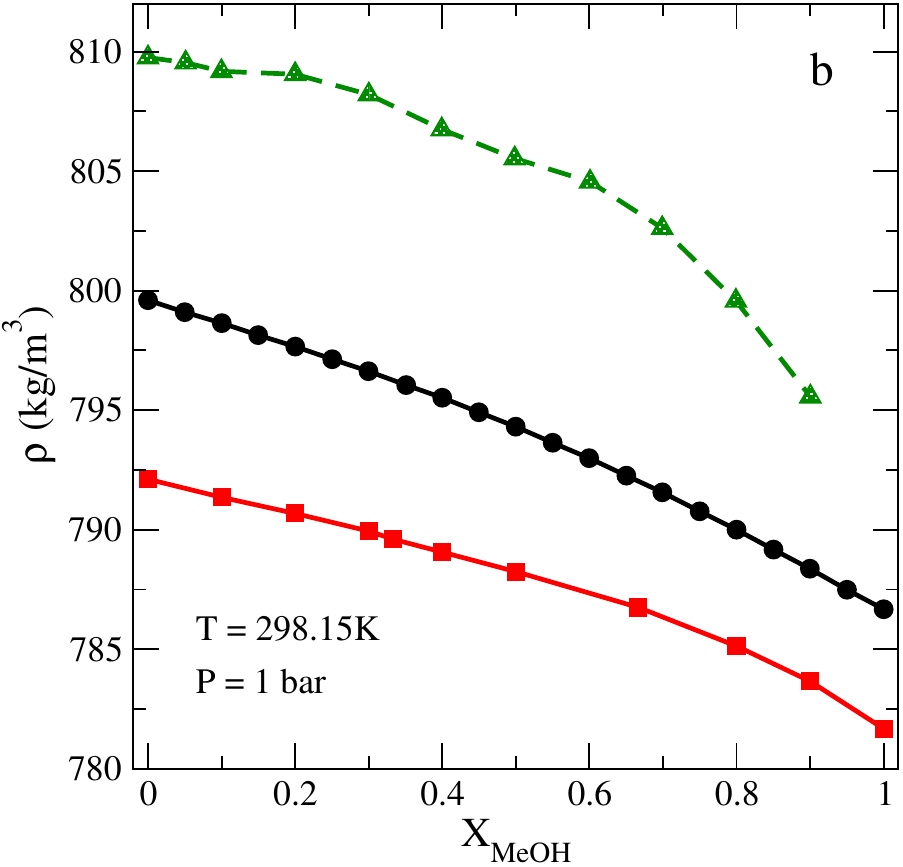}
\end{center}
\caption{(Colour online)
Panel a: a comparison of simulation data (red squares) for the composition dependence of density
for MeOH--PrOH mixture with experimental data (circles) from \cite{kumagai}.
Panel~b:~density of MeOH--PrOH mixture on composition from simulations, 
and experimental data \cite{borun}, nomenclature of symbols as in
panel a, green triangles
reproduced from figure 1a of reference~\cite{madhurima} --- simulation of OPLS/AA model.
}
\label{fig-5}
\protect
\end{figure}

To begin with, a comparison of our simulation data for the composition dependence of
density of MeOH--PrOH solutions with experimental results at different temperature values, 
is shown in figure~\ref{fig-5}. 
It is worth noting that a set of experimental density values from reference~\cite{kumagai}
in figure~\ref{fig-5}a includes solely five points at each of temperatures, $T = 293.15$~K  and $T = 333.15$~K,
in contrast to a more detailed measurements at $T = 298.15$~K from~\cite{borun}.
This work \cite{kumagai} is focused on the study of shear viscosity, rather than
on the precise evaluation of density. Moreover, the experimental point at $X_{\rm MeOH} = 0.755$,
is a bit out of trend, as it can be seen from the lines in figure~\ref{fig-5}a.
The density of mixtures under study smoothly decreases from the value for pure PrOH to
pure MeOH exhibiting a weak non-linearity at each temperature considered.

In general terms, we observe that the UAMI-EW model underestimates the values for density 
in the entire interval of composition
for three temperatures studied, panels a and b of figure~\ref{fig-5}.
However, the magnitude of discrepancy of simulation results
compared to experimental values is of the order of 1\%. Moreover, the shape of $\rho(X_{\rm MeOH})$
dependence from simulations agrees well with experimental trends of behavior of this
property. 

In figure~\ref{fig-5}b, we added the curve (reproduced from reference~\cite{madhurima})
describing the simulation results using the all-atom, OPLS/AA, model for both alcohol species.
The shape of this curve differs from the experimental trend, its non-linearity
is overestimated in comparison with the experimental data. Moreover, at low and 
intermediate methanol concentration, the magnitude of deviation from
experimental results is larger compared with the UAMI-EW united atom model.

\begin{figure}[h]
\begin{center}
\includegraphics[width=6cm,clip]{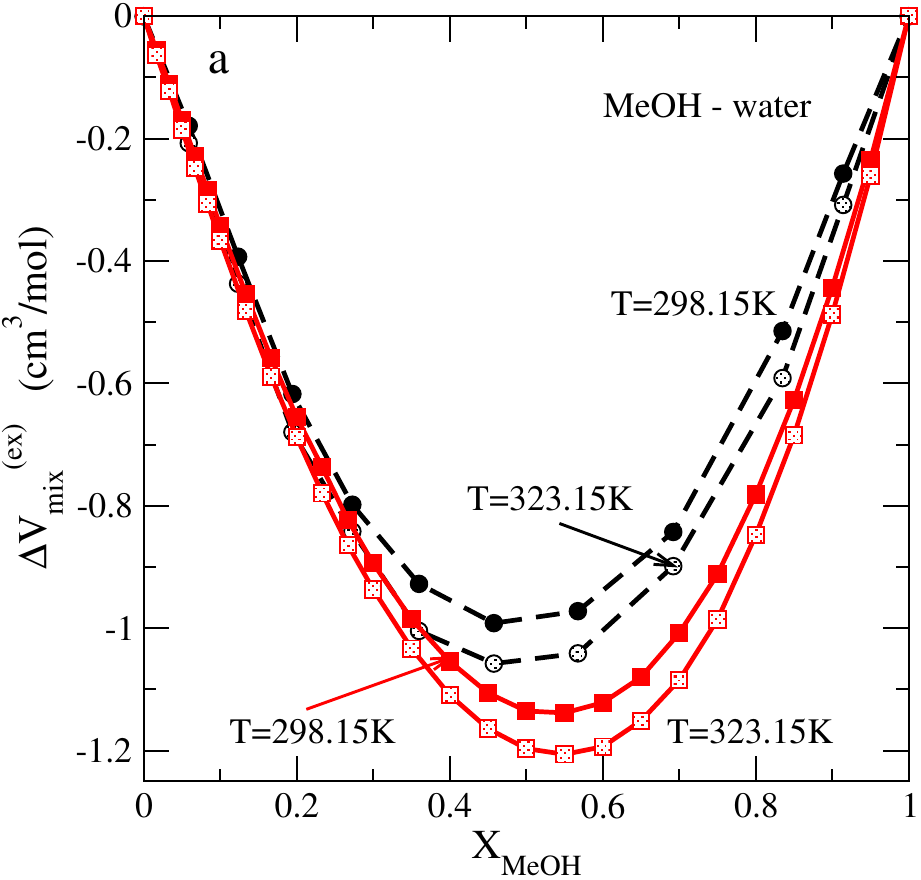}	
\includegraphics[width=6cm,clip]{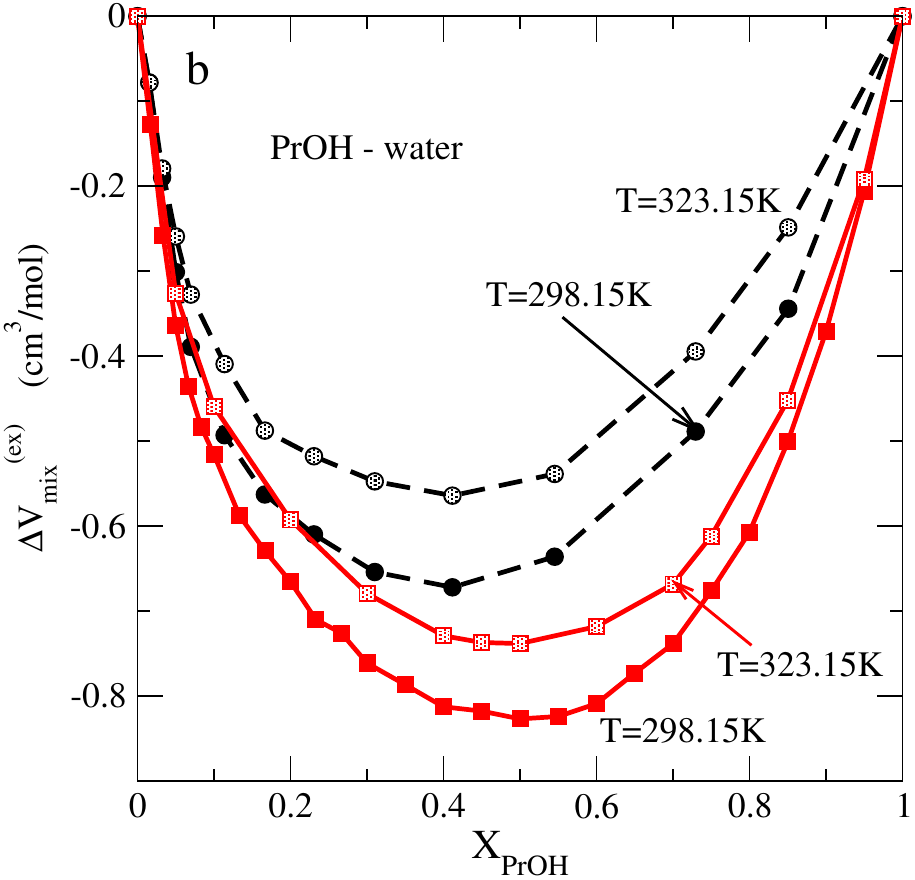}\\
\includegraphics[width=6cm,clip]{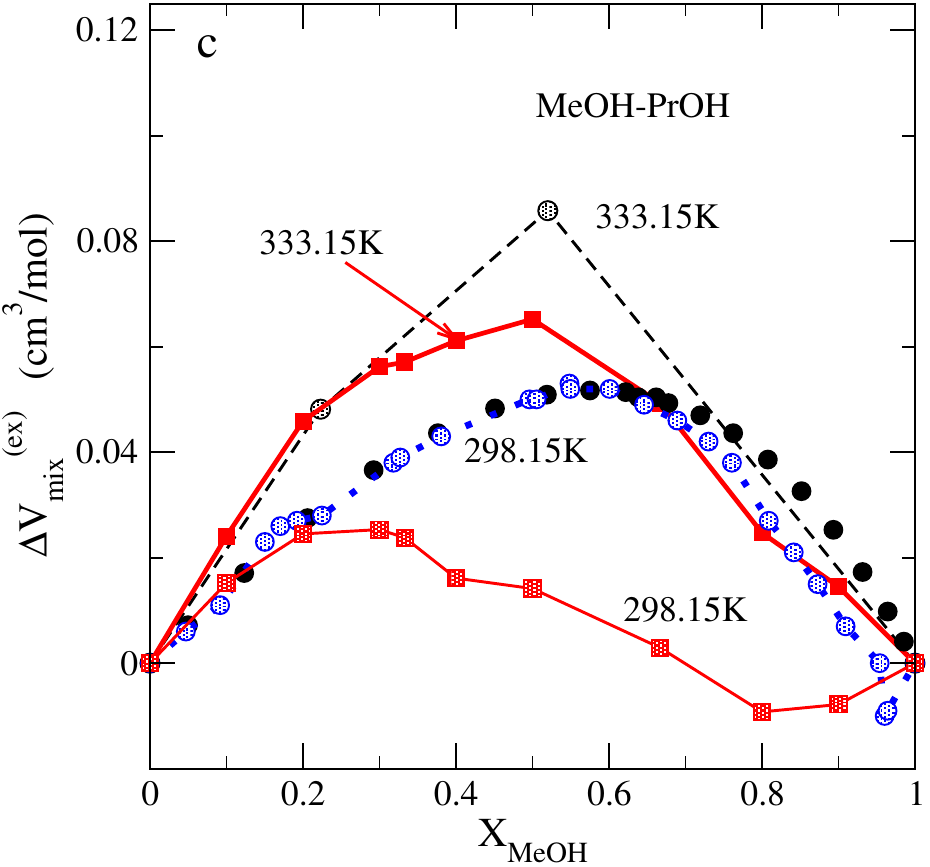}
\end{center}
\caption{(Colour online)
Panels a and b: the excess mixing volume for MeOH--water and PrOH--water mixtures
on composition at two values of temperature, respectively. 
The experimental data (circles) are from Mikhail and Kimmel \cite{mikhail1,mikhail2} 
in both panels, respectively. The simulations are for UAMI-EW model combined 
with TIP4P/$\varepsilon$
model (squares). Panel c:
excess mixing volume for MeOH--PrOH mixture on composition. 
The experimental data are from references~\cite{benson,iglesias} 
at $T=298.15$~K (black and blue circles, respectively). 
The simulations are UAMI-EW model for both alcohols at different temperatures (squares).
The experimental points at $T=333.15$~K (shaded circles) were calculated from the density values
given by Kumagai \cite{kumagai} (the point at $X_{\rm MeOH} = 0.755$ has
been omitted due to inaccuracy).
}
\label{fig-6}
\protect
\end{figure}

Trends of behavior of density upon changes of composition can be interpreted in
terms of the mixing properties. Specifically,
in figure~\ref{fig-6} we plot the dependence of the excess mixing volume,
$\Delta V^{(ex)}_{\rm mix} = V_{\rm mix} - X_{\rm MeOH} V_{\rm MeOH} - (1-X_{\rm MeOH}) V_{\rm PrOH}$,
on $X_{\rm MeOH}$ at different temperatures. 

In order to gain confidence about the 
trends of behavior of $\Delta V^{(ex)}_{\rm mix}$ versus $X_{\rm MeOH}$ upon temperature
change from $T = 298.15$~K to $T = 333.15$~K for a binary alcohol system, 
we performed additional simulations
for MeOH--water and PrOH--water mixtures at $T = 298.15$~K and $T = 323.15$~K
because the experimental data at these conditions are available.
The results are shown in figures~\ref{fig-6}a and~\ref{fig-6}b. The experimental data witness
that the maximum absolute value for $\Delta V^{(ex)}_{\rm mix}$ increases
for MeOH--water mixture upon the temperature change from 298.15~K to 323.15~K (figure~\ref{fig-6}a).
By contrast, opposite behavior is observed for PrOH--water system (figure~\ref{fig-6}b)
within the same temperature interval. The UAMI-EW model for each of two alcohols combined
with the TIP4P/$\varepsilon$ model for water reproduce this behavior 
qualitatively correctly. 
If both alcohols are modelled within the UAMI-EW model, the excess mixing volume
for MeOH--PrOH system behaves as shown in figure~\ref{fig-6}c. 
Two, statistically significant  sets of independent experimental data, 
at $T=298.15$~K, confirm that $\Delta V^{(ex)}_{\rm mix}$
reaches the maximum value at $X_{\rm MeOH} \approx 0.6$. The absolute value for
$\Delta V^{(ex)}_{\rm mix}$ at this particular composition and within the
entire composition range is small, as expected for nearly ideal behavior
of two mixed, quite similar, species. The data set from reference~\cite{iglesias}
predicts negative values for $\Delta V^{(ex)}_{\rm mix}$ 
at $X_{\rm MeOH} \approx 0.95$ (i.e., for systems with very small
amount of 1-propanol dissolved in MeOH), in contrast to dataset from~\cite{benson}
that indicates positive $\Delta V^{(ex)}_{\rm mix}$ in the entire composition range.

The simulation results for UAMI-EW model deviate from the experimental data, 
except for the composition interval corresponding to propanol-rich mixtures. 
Hence, the model is not highly accurate for geometric aspects of mixing of
quite similar alcohol species. Actually, the UAMI-EW model overestimates the influence of PrOH
component in the $\Delta V^{(ex)}_{\rm mix}$ dependence on composition, $X_{\rm MeOH}$.

At $T = 333.15$~K, a comparison of simulation data and experiment is more difficult to perform.
The experimental density values from reference~\cite{kumagai} 
in figure~\ref{fig-5}a include solely five points at $333.15$~K.
These data can be easily converted into the excess mixing volume values to yield three
points only for intermediate $X_{\rm MeOH}$, see, e.g., equation~(2) from reference~\cite{torres}. 
Thus, the results do not provide statistically significant trends. 
Nevertheless, we observe that simulations predict positive values for
$\Delta V^{(ex)}_{\rm mix}$ in the entire composition range in accordance with  experimental values.
In summary, we claim that for binary alcohol mixture in question, 
the excess mixing volume values increase in magnitude upon increasing temperature.
Hence, the observed trend on temperature exhibits similarity to the behavior 
of PrOH--water mixtures shown in figure~\ref{fig-6}b. Unfortunately we are not aware of other studies
of excess mixing volume for binary alcohol mixtures on temperature.

\begin{figure}[h]
\begin{center}
\includegraphics[width=6cm,clip]{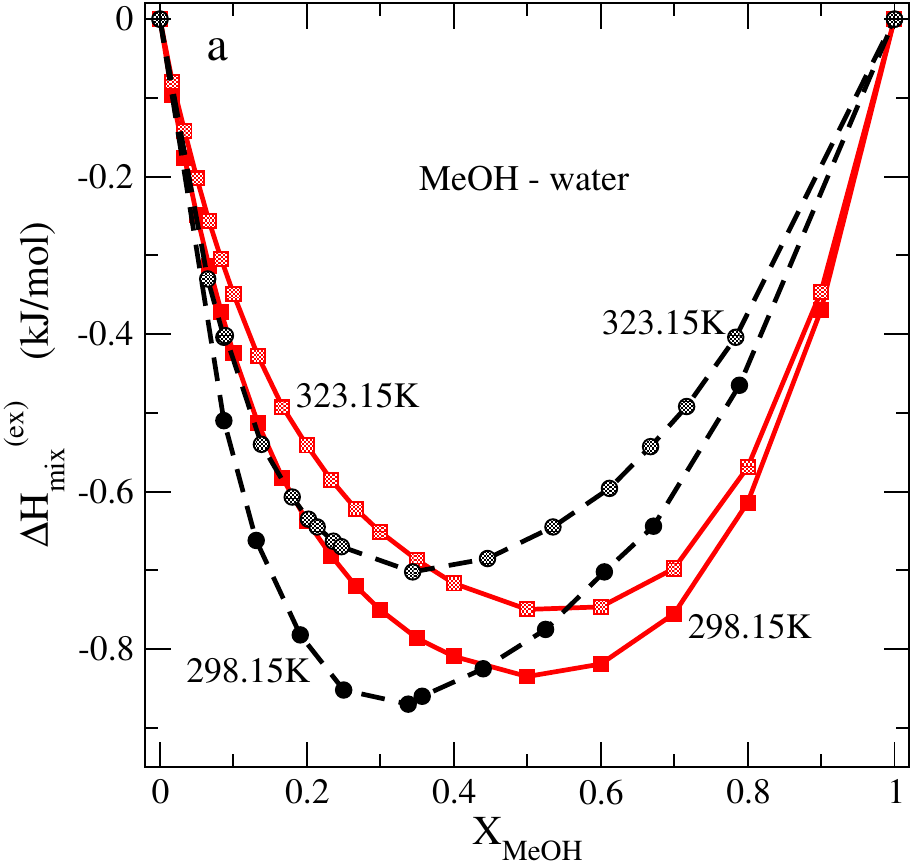}
\includegraphics[width=6cm,clip]{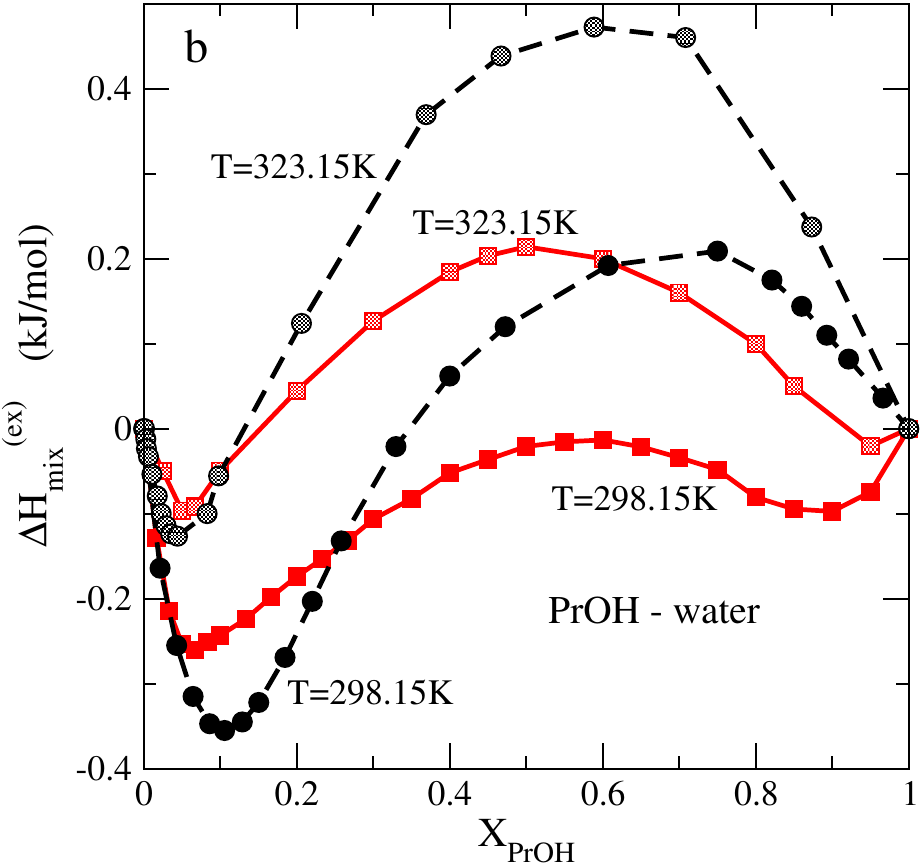} \\
\includegraphics[width=6.5cm,clip]{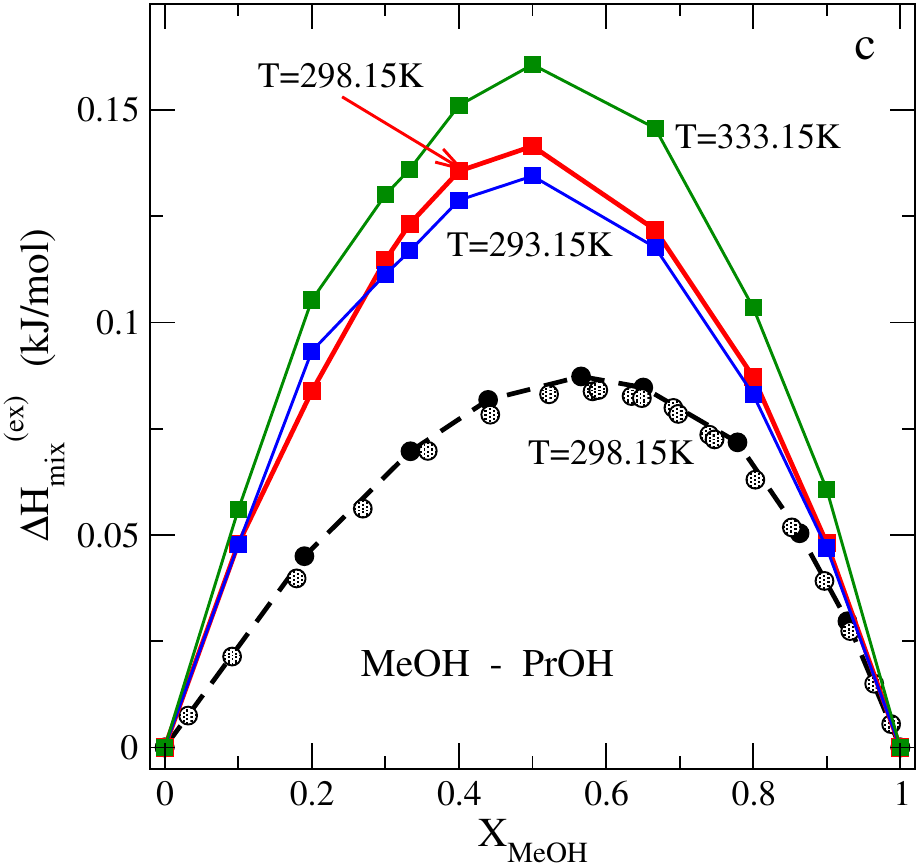}
\end{center}
\caption{(Colour online) 
Panels a and b: the excess mixing enthalpy for MeOH--water and PrOH--water mixtures
on composition at two values of temperature, respectively.
The experimental data in panel a are from reference~\cite{lama,iwona} (solid and shaded black circles, 
respectively)
and from~\cite{marongiu,tsvetov} (solid and shaded circles, respectively)
in panel b.
Panel c: excess mixing enthalpy of methanol--propanol mixture
on methanol mole fraction at 298.15~K and 1 bar. Molecular dynamics simulation results of
UAMI-EW model (red squares). 
Experimental results at 298.15~K are from references~\cite{khurma,benson-heat} (solid
and shaded black circles, respectively).
}
\label{fig-7}
\protect
\end{figure}

Let us proceed now to the exploration of energetic aspects of mixing on temperature.
To do that, we performed a similar analysis as just above for the excess mixing volume. 
The excess mixing enthalpy from simulations and experiments is shown in figure~\ref{fig-7}.
A set of results from additional simulations of alcohol--water mixtures
is given in panels a and b of figure~\ref{fig-7}. The behavior of $\Delta H^{(ex)}_{\rm mix}$
on $X_{\rm MeOH}$ in figure~\ref{fig-7}a for MeOH--water systems witness exothermic mixing of species.
However, the maximum absolute value for $\Delta H^{(ex)}_{\rm mix}$ decreases  upon increasing 
temperature for MeOH--water mixtures. This case was discussed in our
recent publication \cite{mario} more in detail. 

On the other hand, the dependence of $\Delta H^{(ex)}_{\rm mix}$ on $X_{\rm PrOH}$
at different temperature values is more complex (figure~\ref{fig-7}b). The experimental data
for $\Delta H^{(ex)}_{\rm mix}$ at $T = 298.15$~K  and $T = 323.15$~K show
the presence of a minimum of this function for water-rich mixtures and
a maximum for PrOH-rich mixtures. Thus, one observes a transition from
exothermic to endothermic mixing behavior at both temperature values. 
The computer simulation results reproduce temperature trends in agreement
with experimental behavior. However, at each temperature, the composition
trends are only in a qualitative agreement with experiments. The
values for $\Delta H^{(ex)}_{\rm mix}$ from simulations are better reproduced
for water-rich mixtures, figure~\ref{fig-7}b.

The simulation results for binary alcohol mixtures 
overestimate  the magnitude of values for $\Delta H^{(ex)}_{\rm mix}$ in the 
entire interval of compositions, in comparison with experimental data, figure~\ref{fig-7}c. 
In general, the absolute values for $\Delta H^{(ex)}_{\rm mix}$ are smaller
for binary alcohol mixture, in comparison with alcohol--water systems, as actually expected.
The UAMI-EW model correctly predicts the endothermic mixing behavior.
Moreover, it yields the 
symmetric curve with maximum deviation from ideality at $X_{\rm MeOH} = 0.5$.
The experimental data exhibit maximum at a slightly more prevailing
methanol concentration at $T = 298.15$~K. 
On the other hand, 
the maximum of $\Delta H^{(ex)}_{\rm mix}$ from simulations becomes 
larger at $T=333.15$~K than at lower temperatures.
We are not able to verify the 
accuracy of temperature trends because the experimental data are restricted
to $298.15$~K solely.

Concerning the comparison of the present results with findings of other authors,
we would like to attract the attention of the reader to the 
results shown in figure~1a of reference~\cite{siepmann}.
Apparently, the UAMI-EW model predicts the $\Delta H^{(ex)}_{\rm mix}(X_{\rm MeOH})$ dependence
a bit  better than the TraPPE one. At least, the maximum value for 
$\Delta H^{(ex)}_{\rm mix}(X_{\rm MeOH})$ following from UAMI-EW is closer to the 
experimental value.
It is not known if the UAMI-EW force field in its present form is sufficiently good
for primary alcohol--secondary alcohol mixtures. This issue was discussed using TraPPE modelling for the excess mixing enthalpy in reference~\cite{siepmann}.
For UAMI-EW force field, the problem requires a separate study. 

\begin{figure}[h]
\begin{center}
\includegraphics[width=6.5cm,clip]{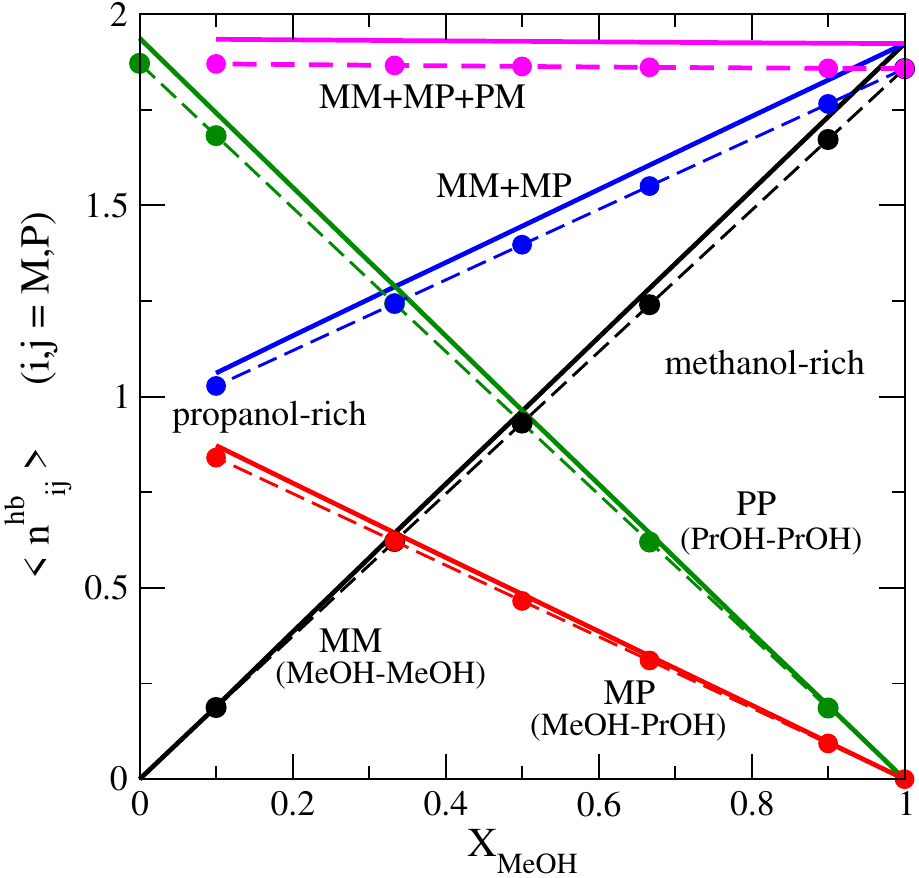}
\end{center}
\caption{(Colour online) 
Average number of hydrogen bonds per molecule in methanol--propanol mixture
on methanol mole fraction. Molecular dynamics simulation results are
for UAMI-EW model for both species. 
Lines are for $T=298.15$~K, whereas the lines decorated with circles are for $T=333.15$~K,
$P = 1$~bar.
}
\label{fig-8}
\protect
\end{figure}

In the recent study of mixtures of primary alcohols and 
of primary and secondary alcohols, Chang and Siepmann~\cite{siepmann} 
discussed the role of hydrogen bonds in modifying the temperature trends 
of mixing enthalpy. Apparently, even small changes of the number of hydrogen bonds
between molecules may lead to a more pronounced changes in the mixing properties.
For the sake of illustration, we show the composition dependence of the average number
of bonds per molecule for MeOH--PrOH mixtures in figure~\ref{fig-8}.
This property was obtained by application of the gmx hbond facility with default
parameters using  GROMACS software.

Each of pure components, MeOH and PrOH, is characterized by a similar value
for the average number of bonds per molecule, $n^{hb}_{MM} \approx n^{hb}_{PP} \approx 2$,
indicating the dominance of chain-like structure.
The lines describing the bonding of species are linear and 
practically symmetric with respect to equimolar composition.
The total number of bonds per molecule of each species is almost constant
in the entire interval of composition, as expected for a mixture of very similar species.
This behavior is in sharp contrast to the changes observed for 
hydrogen bonding in methanol--water mixtures upon composition,
see detailed discussion of this issue in, e.g., reference~\cite{galicia}.
However, the principal conclusion for the system in question, 
is that quite small changes of bonding due to
temperature changes (figure~\ref{fig-8}) are manifested in a more pronounced changes of the excess mixing
enthalpy in figure~\ref{fig-7}c. 

\begin{figure}[h]
\begin{center}
\includegraphics[width=6.0cm,clip]{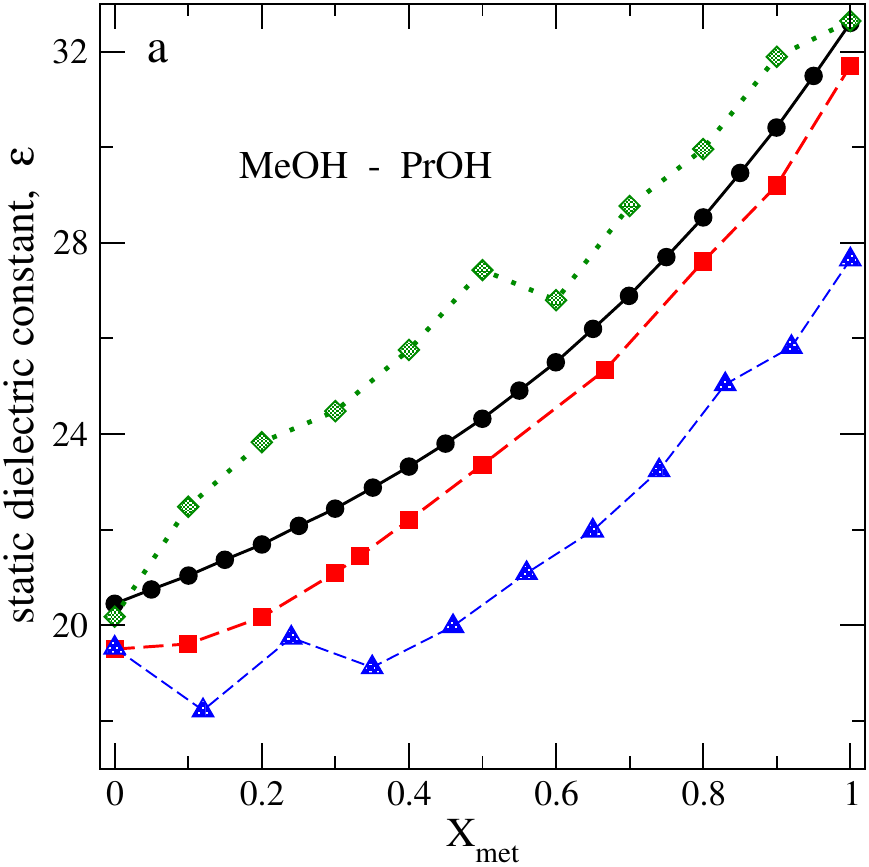}
\includegraphics[width=6.0cm,clip]{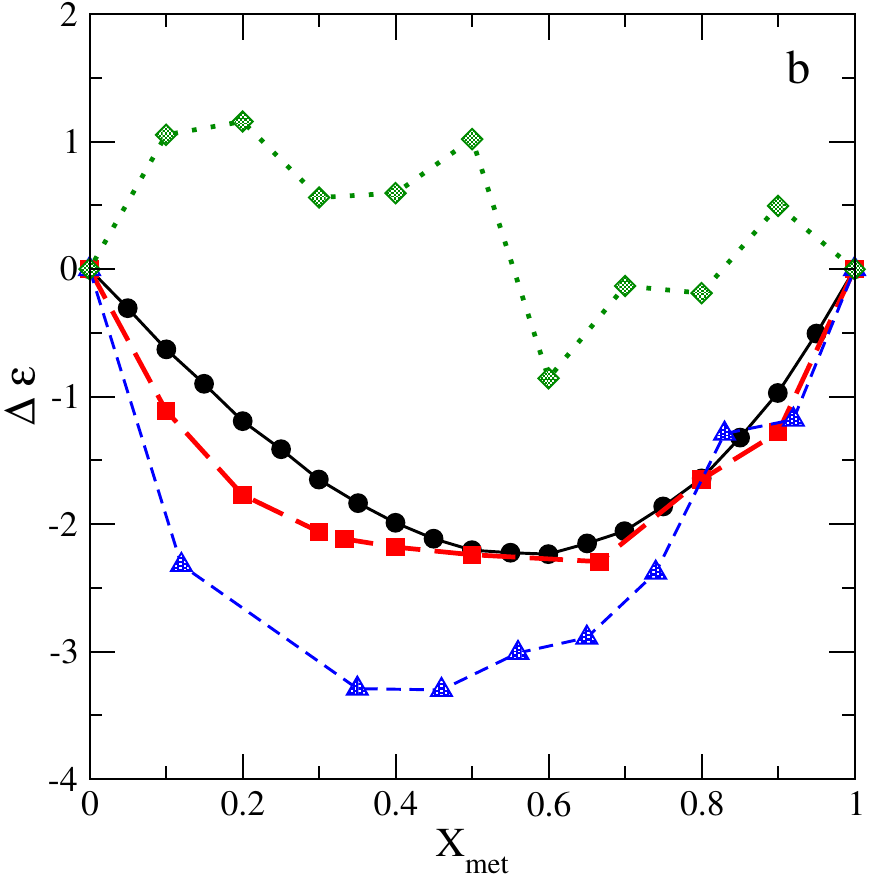}
\end{center}
\caption{(Colour online) 
Panel a: dielectric constant of MeOH--PrOH mixture on mixture composition, 
experiment~\cite{chmielewska} (black solid circles), 
simulation --- UAMI-EW (red squares), ab initio results reproduced from reference~\cite{lone} 
(green diamonds).
Blue triangles --- simulations of OPLS/AA model reproduced from reference~\cite{madhurima}.
Panel b: excess dielectric constant on composition. The nomenclature of lines and
symbols as in panel a.
}
\label{fig-9}
\protect
\end{figure}

Next, we proceed to the comparison of the
dielectric constant and of the excess dielectric constant from our simulations with
experimental data from reference~\cite{chmielewska}. 
Our results are illustrated in
two panels of figure~\ref{fig-9}. From panel a of this figure we learn that the UAMI-EW model
for the MeOH--PrOH mixture pretty well describes the composition behavior of the static
dielectric constant, $\varepsilon (X_{\rm MeOH})$. The model appropriately 
describes the shape of this function. The deviation between simulation results and
experimental data is not big in the entire composition interval. 
By contrast, the differences between the results
of all-atom modelling \cite{madhurima}  and experimental ones are much bigger. On the other hand,
the results from ab initio modelling \cite{lone} also deviate from experimental points.
Inaccuracy of the points on the curve $\varepsilon (X_{\rm MeOH})$ become more pronounced
on the line that shows the excess dielectric constant, figure~\ref{fig-9}b. 
We observe that the UAMI-EW model is very successful in describing the $\Delta \varepsilon (X_{\rm MeOH})$
dependence. A small deviation from the experimental predictions is seen
on the propanol-rich side. The all-atom modelling and ab initio results are much
less successful in the description of $\Delta \varepsilon (X_{\rm MeOH})$ behavior.
Unfortunately, we are not aware of the results for the dielectric constant
of the system in question at temperatures higher and lower than 298.15~K.

\begin{figure}[h]
	\begin{center}
		\includegraphics[width=6.5cm,clip]{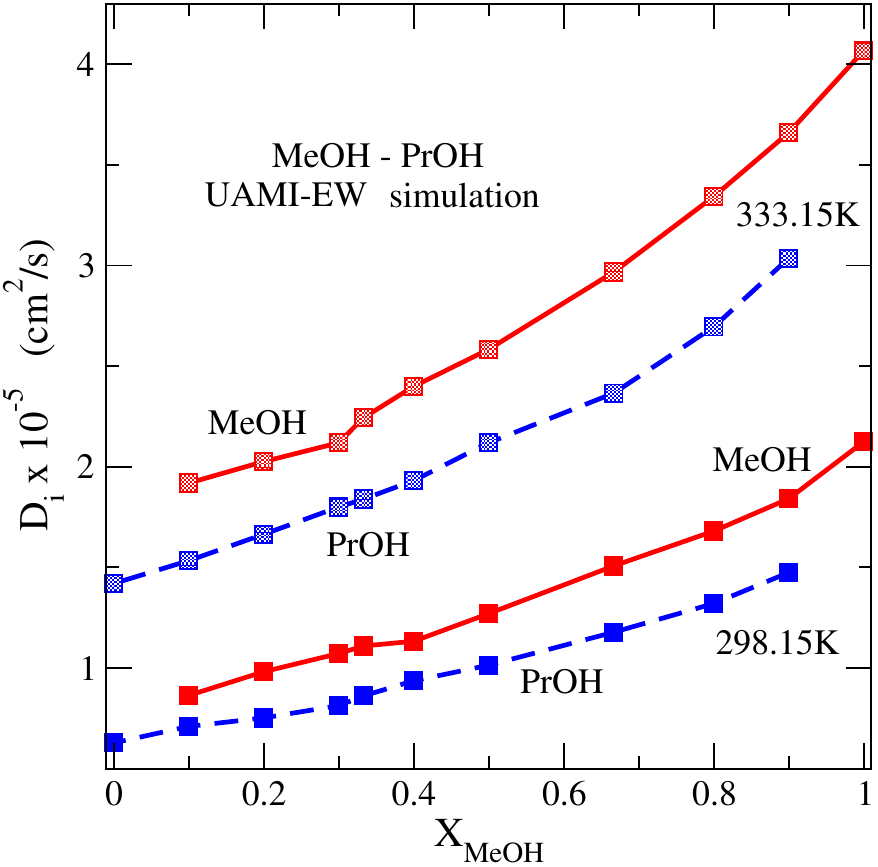}
	\end{center}
	\caption{(Colour online) 
		Self-diffusion coefficients of components in methanol--propanol mixture 
		on methanol mole fraction at $T=298.15$~K and at $T=333.15$~K, at ambient pressure, 1~bar. 
		Molecular dynamics simulation results for UAM-EW model.
	}
	\label{fig-10}
	\protect
\end{figure}

The final issue of our interest in the present work concerns the composition dependence 
of the self-diffusion coefficients of species for the mixtures under study, figure~\ref{fig-10}.
It is worth mentioning that
self-diffusion coefficients may serve as tests of the force fields and 
to seek for their improvement.
The experimental data for self-diffusion coefficients for binary alcohol
mixtures are available for methanol--ethanol mixtures on composition at
298.15~K~\cite{uedaira} (note incorrect spelling of the author's surname, Uedaira,
in the journal web-page, that made tedious our search of the relevant literature). 
Another set of data for MeOH--EtOH system was  published very recently~\cite{golubev}.
In spite of an extensive search, we have not found the data for MeOH--PrOH liquid mixture, unfortunately.

Our simulation results for MeOH--PrOH system, show that the values of self-diffusion coefficients,
$D_i (X_{\rm MeOH})$, for each species increase with increasing methanol amount in the mixture.
$D_{\rm MeOH} (X_{\rm MeOH})$ values are larger than of $D_{\rm PrOH} (X_{\rm MeOH})$, i.e., smaller molecules
move faster than the bigger ones. As smaller inclination of the curves with concentration
of methanol is observed for
propanol-rich mixtures. In qualitative terms, the trends of behavior of $D_i (X_{\rm MeOH})$ 
shown in figure~\ref{fig-10} agree with the experimental observations 
for MeOH--EtOH mixtures, see, e.g., figure~1 of reference~\cite{uedaira} and figure~1 of~\cite{golubev}. There is a small qualitative difference between these two sets 
of experimental data
at a very small methanol concentration that may be attributed to the deuterated
sample used in the former study~\cite{uedaira}. In summary, we are convinced that
the concentration dependence of the self-diffusion coefficients coming from UAMI-EW
simulations is captured satisfactorily.

\section{Concluding remarks}

To conclude, we would like to note that the recently proposed modelling of alcohols within
the UAMI-EW model~\cite{melgarejo} is quite satisfactory for the description of 
properties of MeOH, EtOH and
PrOH (1-propanol) in sufficiently large interval of pressures and of temperature.
The present work complements our recent findings concerning this 
specific modelling~\cite{cmp2026,bermudez,mario}. Several properties, such as density,
excess mixing volume, excess mixing enthalpy, static dielectric constant, surface tension,
self-diffusion coefficients, all of them were critically evaluated and compared with
available experimental data in detail. Some issues concerned the trends of behavior
of a microscopic structure were elucidated as well~\cite{bermudez,mario}. 
In addition, some insights about the changes of the cooperative features 
of hydrogen bonding in methanol upon high pressures have been 
discussed~\cite{jcp2025}.

Rather than enumerating the findings of the present work, 
we prefer to focus on important extensions necessary to deal with. The first of them
in our opinion is to investigate the models left out of attention for the moment.
Namely, a set of more complex alcohols (e.g., 
the 2-propanol, 1- and 2-butanol, see SI of reference~\cite{melgarejo}) 
should be investigated on the same ground
as simpler alcohols of the present study. 
Then, one can attempt to explore mixtures 
of primary alcohols, as well as mixtures of primary and secondary alcohols, similarly
to reference~\cite{siepmann}. Aqueous solutions of such models are necessary to study as well.
Temperature and pressure dependence, besides the composition changes, of various properties 
should be explored. As regards the missing properties necessary to explore, we would like 
just to mention the shear viscosity and dielectric relaxation. Moreover, it is
intersting to provide 
a more detailed description of interfacial phenomena beyond the surface tension.
We expect that our simulation findings  would stimulate the experimental measurements.
Moreover, we expect progress of studies of microscopic structure of alcohols and their
mixtures at high pressures using diffraction techniques, similarly to~\cite{jcp2025}.
At the same time, one may attempt to refine a more sophisticated all-atom type 
models for alcohols and to intend the application of  polarizable models.


\ukrainianpart

\title{II. Температурна залежність властивостей простих одноводневих спиртів. Молекулярно-динамічне моделювання моделі UAMI-EW об'єднаного атома}
%
%
\author{M. Aгілар\refaddr{label1},
	E. Нуньєс-Рохас\refaddr{label2},
	O. Пізіо\refaddr{label1} }

\addresses{
	\addr{label1}
	Інститут хімії, Національний автономний університет Мексики, Сіркуіто Екстеріор, 04510, Мексика
	\addr{label2}
	Факультет хімії, Столичний автономний університет Ізтапалапа,
	просп. Сан Рафаель Атлікско 186, Віцентіна, 09340, Мехіко, Мексика
}

\makeukrtitle

\begin{abstract}
	\tolerance=3000%
	З використанням ізобарно-ізотермічного комп'ютерного моделювання молекулярної динаміки досліджується залежність від температури низки властивостей простих одноводневих спиртів.
	Зокрема, досліджуються метанольний (MeOH), етанольний (EtOH) та 1-пропаноловий (PrOH) спирти.
	З цією метою для кожного зі спиртів застосовується нещодавно запропоноване неполяризовне силове поле об'єднаного атома 
	[V.~Garc\'{i}a-Melgarejo et al., J. Mol. Liq., 2021, \textbf{323}, 114576].
	Точність силового поля обговорюється шляхом порівняння прогнозів моделювання та експериментальних даних для густини, діелектричної проникності, поверхневого натягу, та коефіцієнта самодифузії.
	Додаткові дані щодо застосовності моделі отримані шляхом дослідження залежності різних властивостей від складу сумішей MeOH--PrOH.
	З'ясовано особливості змішування частинок у цій системі з точки зору густини, надлишкового об'єму змішування та надлишкової ентальпії змішування.
	Отримано статичну діелектричну проникність суміші та відповідний надлишок.
	Наприкінці статті прокоментовано перспективи моделювання.
	\keywords метанол, етанол, 1-пропанол, температурна залежність, молекулярно-динамічне моделювання
	
\end{abstract}

\lastpage

\begin{thebibliography}{99}
\bibitem{mathias} Mathias P. M., Ind. Eng. Chem. Res., 2019, {\bf 58}, 12465, \doi{10.1021/acs.iecr.9b01624}.
\bibitem{siepmann} Chang Ch.-K., Siepmann J. I., J. Chem. Eng. Data, 2024, {\bf 69}, 509, \doi{10.1021/acs.jced.3c00415}.
\bibitem{trappe} Chen B., Potoff J. J., Siepmann J. I., J. Phys. Chem. B, 2001, {\bf 105}, 3093, \doi{10.1021/jp003882x}.
\bibitem{melgarejo} Garc\'ia-Melgarejo V., N\'u\~nez-Rojas E., Alejandre J., J. Mol. Liq., 2021, {\bf 323}, 114576,\\ \doi{10.1016/j.molliq.2020.114576}.
\bibitem{cmp2026} Aguilar M., Pusztai L., Pizio O., Condens. Matter Phys., 2026, {\bf 29}, No. 1, 13502, \doi{10.5488/CMP.29.13502}.
\bibitem{jcp2025} Bak\'o I., Pusztai L., Pizio O., J. Chem. Phys., 2025, {\bf 163}, 194504, \doi{10.1063/5.0300069}.
\bibitem{vega-met} Gonzalez-Salgado D., Vega C., J. Chem. Phys., 2016, {\bf 145}, 034508, \doi{10.1063/1.4958320}.
\bibitem{ball} Ball P., Life's Matrix: A Biography of Water, Farrar, Straus, and Giroux, New York, 1999.
\bibitem{franks} Franks F., Water: A Matrix of Life, Royal Society of Chemistry, Cambridge, 2000.
\bibitem{vrhovsek} Vrhov\v{s}ek A., Gereben O., Jamnik A., Pusztai L., J. Phys. Chem. B, 2011, {\bf 115}, 13473, \doi{10.1021/jp206665w}.
\bibitem{weitkamp} Weitkamp T., Neuefeind J., Fischer H. E., Zeidler M. D., Mol. Phys., 2000, {\bf 98}, 125,\\ \doi{10.1080/00268970009483276}.

\bibitem{gromacs} Spoel D., Lindahl E., Hess B., Groenhof G., Mark A. E., Berendsen H. J. C., J. Comput. Chem., 2005, {\bf 118}, 1701, \doi{10.1002/jcc.20291}.

\bibitem{nist} Linstrom~P.~J., Mallard~W.~G. (Eds.), {{NIST Chemistry WebBook}}, {{NIST Standard Reference Database}} 69, {National Institute of Standards and
	Technology}, Gaithersburg MD, 2025, \doi{10.18434/T4D303}.

\bibitem{valtz} Coquelet Ch., Valtz A., Richon D., de la Fuente J. C., Fluid Phase Equilib., 2007, {\bf 259}, 33, \doi{10.1016/j.fluid.2007.04.030}.

\bibitem{sun} Sun T. F., Schouten J. A., Trappeniers N. J., Biswas S. N., J. Chem. Thermodyn., 1988, {\bf 20}, 1089, \doi{10.1016/0021-9614(88)90115-2}.

\bibitem{moreau} Moreau A., Sobrino M., Zambrano J., Segovia J. J., Villama\~{n}an M. A., Mart\'in M. C., J. Mol. Liq., 2021, {\bf 344}, 117744, \doi{10.1016/j.molliq.2021.117744}.

\bibitem{martin} Neumann M., Mol. Phys., 1983, {\bf 50}, 841, \doi{10.1080/00268978300102721}.

\bibitem{kay} Bezman R. D., Casassa E. F., Kay R. L., J. Mol. Liq., 1997, {\bf 73--74}, 397, \doi{10.1016/S0167-7322(97)00082-2}.

\bibitem{dannhauser} Dannhauser W., Bahe L. W., J. Chem. Phys., 1964, {\bf 40}, 3058, \doi{10.1063/1.1724948}.

\bibitem{davidson} Davidson D. W., Can. J. Chem., 1957, {\bf 35}, 458, \doi{10.1139/v57-066}.

\bibitem{moriyoshi} Uosaki Y., Ito K., Kondo M., Kitaura S., Moriyoshi T., J. Chem. Eng. Data, 2006, {\bf 51}, 1915, \doi{10.1021/je060248p}.

\bibitem{strey} Strey R., Schmeling T., Ber. Bunsen Ges. Phys. Chem., 1983, {\bf 87}, 324, \doi{10.1002/bbpc.19830870411}.

\bibitem{souckova} Sou\v{c}kova M., Klomfar J., P\'atek J., J. Chem. Eng. Data, 2008, {\bf 53}, 2233, \doi{10.1021/je8003468}.

\bibitem{vazquez} V\'azquez G., Alvarez E., Navaza J. M., J. Chem. Eng. Data, 1995, {\bf 40}, 611, \doi{10.1021/je00019a016}.

\bibitem{hurle} Hurle R. L., Easteal A. J., Woolf L. A., J. Chem. Soc., Faraday Trans. 1, 1985, {\bf 81}, 769, \doi{10.1039/F19858100769}.

\bibitem{karger} Karger N., Vardag T., L\"udemann H.-D., J. Chem. Phys., 1990, {\bf 93}, 3437, \doi{10.1063/1.458825}.

\bibitem{pratt} Pratt K. C., Wakeham W. A., J. Chem. Soc., Faraday Trans. 2, 1977, {\bf 73}, 997, \doi{10.1039/F29777300997}.

\bibitem{shaker} Shaker-Gaafar N., Karger N., Wappmann S., L\"udemann H.-D., Ber. Bunsen Ges. Phys. Chem., 1993, {\bf 97}, 805, \doi{10.1002/bbpc.19930970610}.

\bibitem{bermudez} Mendez-Bermudez J. G., Pizio O., J. Mol. Liq., 2025, {\bf 421}, 126789, \doi{10.1016/j.molliq.2024.126789}.

\bibitem{wensink} Wensink E. J. W., Hoffmann A. C., van Maaren P. J., van der Spoel D., J. Chem. Phys., 2003, {\bf 119}, 7308, \doi{10.1063/1.1607918}.

\bibitem{anisimov} Anisimov V. M., Vorobyov I. V., Roux B., MacKerell A. D., J. Chem. Theory Comput., 2007, {\bf 3}, 1927, \doi{10.1021/ct700100a}.

\bibitem{madhurima} Madhurima V., Ajmal Rahman M. K., Saishree K., Abdulkareem U., Phys. Chem. Liq., 2024, {\bf 62}, 9, \doi{10.1080/00319104.2023.2263897}.

\bibitem{lone} Lone B., Madhurima V., J. Mol. Model., 2011, {\bf 17}, 709, \doi{10.1007/s00894-010-0772-y}.

\bibitem{kumagai} Kumagai A., Yokoyama C., Int. J. Thermophys., 1998, {\bf 19}, 3, \doi{10.1023/A:1021438800094}.

\bibitem{borun} Boru\'n A., \'Zurada M., Bald A., J. Therm. Anal. Calorim., 2010, {\bf 100}, 707, \doi{10.1007/s10973-009-0157-6}.

\bibitem{mikhail1} Mikhail S. Z., Kimel W. R., J. Chem. Eng. Data, 1961, {\bf 6}, 533, \doi{10.1021/je60011a015}.

\bibitem{mikhail2} Mikhail S. Z., Kimel W. R., J. Chem. Eng. Data, 1963, {\bf 8}, 323, \doi{10.1021/je60018a014}.

\bibitem{benson} Benson G. C., Pflug H. D., J. Chem. Eng. Data, 1970, {\bf 15}, 382, \doi{10.1021/je60046a020}.

\bibitem{iglesias} Iglesias M., Orge B., Tojo J., J. Chem. Eng. Data, 1996, {\bf 41}, 218, \doi{10.1021/je950186v}.

\bibitem{torres} Torres R. B., Marchiore A., Volpe P., J. Chem. Thermodyn., 2006, {\bf 38}, 526, \doi{10.1016/j.jct.2005.07.012}.

\bibitem{lama} Lama R. F., Lu B. C.-Y., J. Chem. Eng. Data, 1965, {\bf 10}, 216, \doi{10.1021/je60026a003}.

\bibitem{iwona} Tomaszkiewicz I., Randzio S. L., Gierycz P., Thermochim. Acta, 1986, {\bf 103}, 281, \doi{10.1016/0040-6031(86)85164-4}.

\bibitem{mario} Cruz Sanchez M., Trejos Montoya V., Pizio O., Condens. Matter Phys., 2025, {\bf 28}, 13602, \doi{10.5488/CMP.28.13602}.

\bibitem{marongiu} Marongiu B., Ferino I., Monaci R., Solinas V., Torrazza S., J. Mol. Liq., 1984, {\bf 28}, 229, \doi{10.1016/0167-7322(84)80027-6}.

\bibitem{tsvetov} Tsvetov N., Sadaeva A., Toikka M., Zvereva I., Toikka A., J. Therm. Anal. Calorim., 2020, {\bf 142}, 1977, \doi{10.1007/s10973-020-09605-y}.

\bibitem{khurma} Khurma J. R., Fenby D. V., J. Phys. Chem., 1979, {\bf 83}, 2443, \doi{10.1021/j100482a004}.

\bibitem{benson-heat} Pflug H. D., Pope A. E., Benson G. C., J. Chem. Eng. Data, 1968, {\bf 13}, 408, \doi{10.1021/je60038a032}.

\bibitem{galicia} Galicia-Andr\'es E., Dominguez H., Pusztai L., Pizio O., J. Mol. Liq., 2015, {\bf 212}, 70, \doi{10.1016/j.molliq.2015.08.061}.

\bibitem{chmielewska} Chmielewska A., \v{Z}urada M., Klimaszewski K., Bald A., J. Chem. Eng. Data, 2009, {\bf 54}, 801, \doi{10.1021/je800593p}.

\bibitem{uedaira} Eudaira H., Inorg. Chim. Acta, 1980, {\bf 40}, X104, \doi{10.1016/S0020-1693(00)92195-9}.

\bibitem{golubev} Golubev V. A., J. Mol. Liq., 2024, {\bf 414}, 126266, \doi{10.1016/j.molliq.2024.126266}.






\end{thebibliography}
\end{document}